\documentclass[twocolumn]{aastex631}

\def\rev#1{#1\rm}
\def\reveng#1{#1\rm}
\usepackage{amsmath}
\usepackage{float}
\usepackage{accents}
\usepackage{comment}

\makeatletter
\AtBeginDocument{%
  \@ifundefined{thelinenumber}{}{%
  }%
  \@ifundefined{makeLineNumber}{}{%
  }%
}
\makeatother

\shorttitle{Metallic Iron in White Dwarf Disks}
\shortauthors{Okuya et al.}

\begin{document}

\title{A Possible Indication of Metallic Iron in White Dwarf Dusty Disks from their ``Dirtiness''}

\correspondingauthor{Ayaka Okuya}
\email{ayaka.okuya.astro@gmail.com}

\author[0000-0001-6222-9423]{Ayaka Okuya}
\affiliation{Department of Physics, University of Warwick, Coventry CV4 7AL, UK}

\author[0000-0002-1886-0880]{Satoshi Okuzumi}
\affiliation{Department of Earth and Planetary Sciences, Institute of Science Tokyo, 2-12-1 Ookayama, Meguro-ku, Tokyo 152-8551, Japan}

\author[0000-0002-0649-6997]{Aki Takigawa}
\affiliation{Department of Earth and Planetary Science, The University of Tokyo,7-3-1 Hongo, Tokyo 113-0033, Japan}

\author[0009-0005-4508-0723]{Hanako Enomoto}
\affiliation{Department of Earth and Planetary Science, The University of Tokyo,7-3-1 Hongo, Tokyo 113-0033, Japan}



\begin{abstract}
Polluted white dwarfs provide unique constraints on the elemental compositions of planetary bodies. The tidal disruption of accreting bodies is thought to form circumstellar dusty disks, whose emission spectra could offer additional insights into the mineral phases of the accreted solid material. Silicates are detected in the mid-infrared spectra of several disks, but do not fully account for the near-infrared excess in the disks' spectra. Conductive materials, such as metallic iron, are potential sources of near-infrared emissivity.
We investigate the role of metallic iron within silicate dust in the observed spectra of the white dwarfs G29-38 and GD56.
Using thermal emission spectra calculations, we analyze the abundance of metallic iron in the dust and the disk structure parameters that best fit the observed spectra.
We find that \rev{metallic-iron-bearing} dust enhances the near-infrared opacity, thereby providing a better fit to the G29-38 spectrum \rev{for various silicate compositions than metallic-iron-free} dust.
The best-fit metal-to-silicate mixing ratio is approximately unity, and for Mg-rich pyroxenes, this value is also consistent with G29-38's stellar atmospheric composition within 1-$\sigma$ observational uncertainties.
\rev{Based on the spectral fitting and compositional consistency, Fe-rich silicates without metallic iron cannot be ruled out.}
The observed GD56 spectrum also favors iron-bearing dust. However, the large observational uncertainties of GD56's stellar elemental abundances hinder a precise comparison between the stellar and dust iron abundances. Upcoming high-precision JWST observations will provide a larger sample, enabling statistical analysis of the correlation between the iron abundances in the atmospheres and circumstellar dust of polluted white dwarfs.




\end{abstract}



\section{Introduction} \label{sec:intro}
White dwarfs, which exhibit signatures of accreted planetary materials in their atmospheres, are valuable targets for investigating the elemental composition of extrasolar planetary objects \citep[e.g.,][]{Zuckerman+2003,Zuckerman+2010, Koester+2014,Hollands+2017}.
Metals (elements heavier than H/He) in a white dwarf atmosphere gravitationally settle within a timescale of Myrs \citep{Paquette1986}, but their common detections beyond this timescale suggest external sources of the pollution \citep{Zuckerman+2003}.
As elemental abundance patterns observed in white dwarf atmospheres resemble that of the solar system planetary bodies, the metals are believed to originate from planets and asteroids that survived the post-main-sequence evolution \citep{Jura&Young2014, Farihi2016, Veras2021}.

A circumstellar debris disk is believed to form through the tidal disruption of the accreting planetary bodies, which release fragments into the white dwarf's surroundings \citep{Jura2003,Veras+2014, Veras+2015,Malamud+2020a,Malamud+2020b,Li+2021,Brouwers+2022}. Dust generated by mutual collisions of the fragments would eventually evaporate into gas and accrete onto the star \citep{Rafikov2011,Kenyon+2017,Okuya+2023}. 
Consequently, the dust composition in disks can provide clues to the solid constituents of the planetary bodies before they break down into atomic/ionic forms detected in the stellar atmosphere.


\rev{Over 200 white-dwarf disks have been identified via their infrared excesses}
\citep{Zuckerman+1987,Jura+2007, Rocchetto+2015}, and previous infrared spectroscopic observations have revealed the composition of the dust in some of these disks. Infrared spectroscopy using Spitzer IRS detected 10-$\mu$m silicate features in 8 systems so far \citep{Reach+2005,Jura+2007AJ,Jura+2009AJ}.These systems exhibit a 12-$\mu$m red wing, suggesting the presence of olivine-rich silicates \citep{Jura+2009AJ}\footnote{\rev{
Their spectral fitting used the \citet{Dorschner+1995} complex refrative index data, whose silicate feature may be affected by Fe$^{3+}$ contamination (see Sections~\ref{subsec:opacity-model} and~\ref{subsec:fe-rich-silicate} for details).
}}. 
\rev{Alternatively, the 12-$\mu$m wing may be related to Al--O bonds, either in alumina grains or Al-bearing silicates.}
The absence of PAH emission features at 3.3 $\mu$m and 5-15 $\mu$m suggest that the disks may be carbon-poor, which aligns with observations of white dwarf atmospheres \citep{Jura+2009AJ,Jura&Young2014}.

Further advancements in the study of the dust composition are expected with the spectroscopic observations using the James Webb Space Telescope (JWST). JWST observations using its Mid-infrared Instrument (MIRI) have already been conducted on systems where silicate features were previously detected by Spitzer \citep{Kate+2023jwst, Reach+2025}. Additionally, a tentative detection of carbonate features at a wavelength of 7 $\mu$m has been reported \citep{Swan+2024}. 
Moreover, a white-dwarf disk survey using JWST has been carried out in Cycle 2 \rev{and has recently revealed diverse silicate features in 12 newly detected disks, increasing the detection rate of debris disks by a factor of about three compared to Spitzer \citep{Farihi+2025}.}

However, an optically thin disk composed solely of silicate dust is known to fail to account for the near-infrared emission in the observed spectra of several white dwarf systems. For example, \citet{Reach+2009} attempted to explain the near-infrared excess in the observed spectrum of G29-38 by adding amorphous carbon to the disk, which has high opacity at near-infrared wavelengths. However, their calculated spectrum was not fully self-consistent as it did not account for the disk structure. In contrast, \citet{Ballering+2022} performed radiative transfer calculations that considered the disk structure but used the ``astronomical silicate'' \citep{Draine&Lee1984}, a hypothetical material with assumed optical properties (see below).
In addition to G29-38, \citet{Jura+2007AJ} and \citet{Jura+2009AJ} attempted to reproduce the observed spectra of other systems by modifying the disk structure instead of the dust composition. They proposed a model with two radial disk components, but it has not yet been rigorously validated.

In the 1970s, it was already recognized that pure \rev{magnesium} silicates produced in laboratory experiments is too transparent to reproduce the observed near-infrared emission from circumstellar and interstellar dust grains. To address this limitation, the concept of the hypothetical material, ``dirty silicate'' was introduced \citep{Jones&Merrill1976}. 
Its infrared dielectric function is derived based on infrared emissivities inferred from astronomical observations. 
The aforementioned astronomical silicate of \citet{Draine&Lee1984} is an example of dirty silicates.
Using an effective-medium theory, \citet{Ossenkopf+1992} demonstrated that potential contributors to the near-infrared opacity are metallic iron, iron oxides, and amorphous carbon with high refractive indices. 
\citet{Kemper+2002}, \citet{Jager+2003}, \rev{and \citet{Speck+2015}} also demonstrated that metallic iron inclusions significantly enhance the near-infrared opacity of silicates.
These conducting materials can absorb electromagnetic waves across a broader wavelength range due to the presence of free electrons.

In this paper, we propose that conducting materials such as metallic iron mixed into silicate dust could be responsible for the near-infrared emission in the observed spectra of two white dwarfs. To test this hypothesis, we calculate the thermal emission spectra of dust with varying its metallic iron abundances, incorporating the disk structure, and fit the results to the observed spectra.
Our analysis shows that dust containing metallic iron or iron oxides provides a better match to the observed spectra than dust without them. Furthermore, the best-fit composition of the iron-bearing dust is consistent with the elemental abundances inferred from the white dwarf atmospheres, within observational uncertainties.


The organization of this paper is as follows. In Section \ref{sec:method}, we describe the observational data used in our analysis, the calculation method for mixed-material dust opacity, and the disk thermal emission model. Section \ref{sec:results} presents theoretical infrared spectra of dust with and without metallic iron, demonstrating that an addition of metallic iron better reproduces both the near- and mid-infrared spectra of G29-38 and GD56, respectively. 
In Section \ref{sec:discussion}, we discuss the consistency between the iron abundance in circumstellar dust and stellar atmospheres, other material candidates contributing near-infrared emission, and the implications of iron's presence in the dust. We summarize our findings in Section \ref{sec:conclusion}.

\section{Method} \label{sec:method}
Through the spectral analysis, we examine the near-infrared emission primarily originates from the dust composition rather than disk physical parameters.
In Section \ref{subsec:data}, we describe the observational data of two white dwarfs, G29-38 and GD56, used in our analysis. In Section \ref{subsec:dustcomp}, we explain the dust compositions assumed in our spectral fitting. Section \ref{subsec:opacity-model} focuses on the calculation methods for the assumed dust opacities, with particular attention to the differences in opacity between conductors and insulators.
We describe our disk emission model and the range of disk parameters (Section~\ref{subsec:disk-model}) and the fitting procedures (Section~\ref{subsec:fitting}).

\subsection{Observational Data} \label{subsec:data}
We select G29-38 and GD56 as the targets for analyzing disk dust composition for the following reasons. First, the near- to mid-infrared spectra of both systems, provided by Spitzer, are available \citep{Reach+2009, Jura+2009AJ}. Second, the elemental abundances of major rock-forming elements, namely Fe, Mg, and Si, in the atmospheres of the white dwarfs have been measured. Finally, both white dwarfs are DA-type with short sedimentation times, making it reasonable to expect that their atmospheric compositions reflect the dust compositions in the surrounding disks.
We note that LTT8452 also satisfies these conditions, but we exclude it from our analysis due to issues with point-source overexposure \citep{Jura+2009AJ}.

The stellar parameters of G29-38 and GD56 are listed in Table \ref{tab:stellar-p}.
We determine the stellar radius of GD56 using $R_{\star}/d$ from the J and K band fitting in \citet{Jura+2009AJ}. The value of $R_{\star}$ predicted by atmospheric structure model is 1.4 times larger than this value. 
We have checked that using this larger $R_{\star}$ in spectral fitting would only affect the best-fitting disk physical parameter but not affect the best-fitting dust composition.
We calculate the flux density of white dwarfs assuming black body radiation at $T_{\star}$.

\begin{table}
\centering
 \caption{Stellar \rev{Parameters} \label{tab:stellar-p}}
 \begin{tabular}{ccccc}
  \hline  \hline
  WD name & $T_{\star}$ (K) & $\log g$ & $R_{\star}$ (1$0^8$ cm)  & $d$ (pc) \\
  \hline
  G29-38$^1$ & 11240 & $8.00$ & $9.04$ & $17.5$   \\
GD56 & 14400$^2$ &  $7.80^2$ & 
 $7.51^3$ & $71.5^4$ \\
  \hline 
 \end{tabular}
 \tablecomments{
  References (1) \citet{Subasavage+2017}; (2) \citet{Koester&Wilken2006}; (3)\citet{SIMBAD}; (4)
  \citet{Jura+2009AJ} and also see the text}
\end{table}

\begin{figure*}
\centering
    \includegraphics[bb=0 0 360 252, width=1\columnwidth]{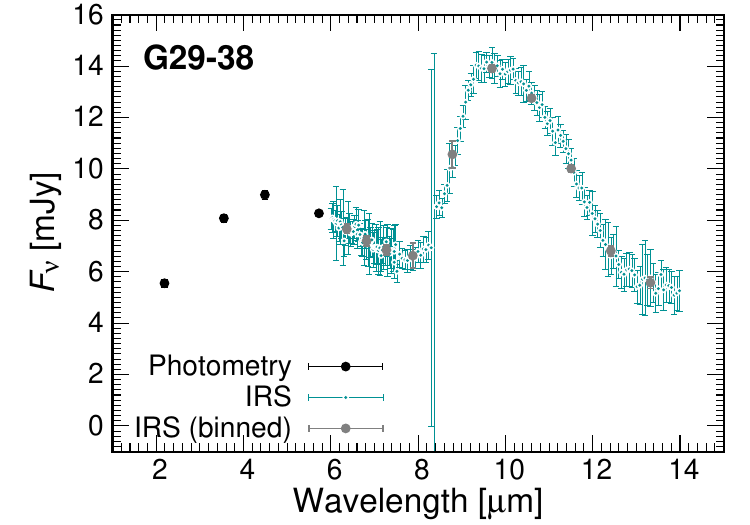}
    \includegraphics[bb = 0 0 360 252, width=1\columnwidth]{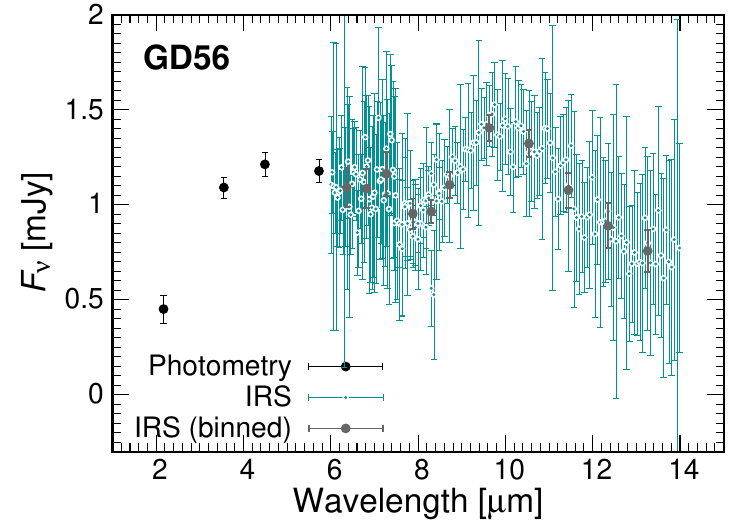}
\caption{Infrared spectra of the G29-38 and GD56 systems (left and right panels, respectively) used in this study. The Spitzer IRS spectra, binned Spitzer IRS spectra, and photometric data are plotted in green, gray, and black, respectively. The error bars represent 1$\sigma$ observational noises. 
}
\label{fig:binning}
\end{figure*}

We use spectra from the Spitzer Space Telescope IRS (\rev{$5.2-14.5~\mu$m}, SL2 and SL1 modules) and photometric data from Spitzer IRAC (3.6 $\mu$m, 4.5 $\mu$m, 5.7 $\mu$m) and the 2MASS K-band (2.2 $\mu$m).
The IRS spectra are obtained from the Combined Atlas of Sources with Spitzer IRS Spectra \citep{Lebouteiller+2011} online database. For G29-38 and GD56, we use AOR 22957568 from 2007 and AOR 22844672 from 2007-10-05, respectively. We utilize the data points at $> 6 \mu$m and group 15 points into one bin (Figure \ref{fig:binning}). The bins have wavelength spacings similar to those of the four photometric data points. We assume the noise follows Poisson statics, for simplicity.
For IRAC photometric data, multiple epochs of data are available for G29-38. Following  \citet{Ballering+2022}, we use their median value. 
The 2MASS Ks-band photometry data for G29-38 and GD56 are provided by \citet{Cutri+2003} and  \citet{Kilic+2006}, respectively.

It should be noted that the observed spectra of G29-38 and GD56 exhibit temporal variations. 
In particular, the 3.6 $\mu$m and 4.5 $\mu$m band data for GD56 show \rev{episodic flux} variations of up to 30\% \rev{on a timescale of $\sim$5~years.} 
For GD56, we use the data taken on 2006-09-20 \citep{Jura+2007}, as it is the only available data at 5.7 $\mu$m \rev{and was obtained close in time to the aforementioned Spitzer/IRS observations, compared to the variation timescale, so the flux levels are indeed similar.
Furthermore, no significant color change has been reported for the GD56 disk, 
indicating that the variability is likely caused by changes in the emitting disk surface area due to episodic dust production rather than compositional changes \citep{Farihi+2018}. 
Therefore, this variability is not expected to significantly affect our compositional analysis. }
In other cases, the flux variations are generally within $\sim$6\%, and hence the choice of the data is not expected to
significantly affect the results of our analysis.

\subsection{Assumed Dust Composition} \label{subsec:dustcomp}

\begin{table*}
\centering
 \caption{Measured Atmospheric Abundances \label{tab:elements}}
 \begin{tabular}{cccccc}
  \hline  \hline
  WD name & C & O & Mg & Si  & Fe\\
  \hline
  G29-38$^1$ & $-6.90 \pm 0.12$ & $-5.00 \pm 0.12$ &
 $-5.77 \pm 0.13$ & $-5.60 \pm 0.17$ & 
$-5.90 \pm 0.10$  \\
GD56$^2$ & -- &  -- & 
 $-5.55 \pm 0.20$ & $-5.69 \pm 0.20$ &
 $-5.44 \pm 0.20$ \\
  \hline 
 \end{tabular}
 \tablecomments{
  The values are given in log unit.
  References (1) \citet{Xu+2014}; (2) \citet{Xu+2019}
  }
\end{table*}
We infer candidate dust compositions based on the measured elemental abundances in the white dwarf atmospheres (Table \ref{tab:elements}), taking into account the  sinking timescale ($t_{\rm sink}$) for each element. The $t_{\rm sink}$ values are obtained from the Montreal White Dwarf Database, where we input $T_{\star}$ and $\log g$ provided in Table~\ref{tab:stellar-p}.
The metal-sinking timescale is $\sim10$ years for G29-38 and $\sim 0.01$ years for GD56. Since these timescales are much shorter than the observationally estimated disk lifetimes \citep{Girven+2012}, we can assume that the stellar abundances are in steady state, and the relative elemental composition of the pollutant  can be calculated as \citep{Koester2009},
\begin{equation}
     \frac{n_{\rm pollutant, A}}{n_{\rm pollutant, B}} = \biggl(\frac{n_{\rm star, A}/t_{\rm sink, A}}{n_{\rm star, B}/t_{\rm sink, B}}  \biggr),
\end{equation}
where $n$ is the number abundance of each element.
The resulting pollutant elemental compositions are summarized in Table~\ref{tab:ss-elements}.
\begin{table}
\centering
 \caption{Steady-state Pollutant Elemental Abundances \label{tab:ss-elements}}
 \begin{tabular}{ccccc}
  \hline \hline
  WD name & C/Si & O/Si & Mg/Si & Fe/Si\\
  \hline
  G29-38 & 0.03$^{+0.02}_{-0.01}$ & 
3.2$^{+2.5}_{-1.8}$ &  0.8$^{+0.7}_{-0.5}$  & 0.7$^{+0.6}_{-0.4}$  \\
GD56 & -- &  -- & 
0.9$^{+1.0}_{-0.7}$ & 3.2$^{+3.7}_{-2.4}$  \\
  \hline 
\end{tabular}
\end{table}

We assume that the dust in the accreting disk consists of silicate, metallic iron, and amorphous carbon. 
For simplicity, each model includes silicate of only a single chemical composition, but we vary the composition to study how it affects the spectral fitting.   
We only consider amorphous silicate as it dominates the silicate features of the observed spectrum of G29-38.
The nominal dust composition in our model is given by
\begin{align}
&{\rm Mg}_x{\rm Fe}_{(2-x)}{\rm SiO}_4 + y {\rm Fe} + z{\rm C}, \quad x = 0.4, 0.5, 1, 2, \label{eq:y-olivine} \\
 &{\rm Mg}_x{\rm Fe}_{(1-x)}{\rm SiO}_3 + y {\rm Fe} + z{\rm C}, \quad x =  0.5, 0.7, 1,  \label{eq:y-pyroxene}
\end{align}
where $y$ and $z$ are free parameters.
Equations \eqref{eq:y-olivine} and \eqref{eq:y-pyroxene} apply to olivines and pyroxenes, respectively.
The Mg/Si and O/Si ratios in the dust composition range from 0.4 to 2 ($x=0.4-2$) and 3 to 4, respectively, consistent with the pollutant elemental abundance of the star shown in Table \ref{tab:elements} within the 1$\sigma$ observational uncertainties.

\rev{
Although the lack of PAH features suggests that the disk dust is carbon-poor (see Section~\ref{sec:intro}), 
we still include a minimal amount of amorphous carbon to ensure consistency with the photospheric C abundance (Table~\ref{tab:ss-elements}). 
As amorphous carbon is a potential contributor to the near-infrared opacity, 
we evaluate its impact on the model spectrum and show that this small amount has a negligible effect on the spectral fit (to be discussed in  Section~\ref{subsec:other-conductor}). 
Regarding the carbon-bearing phase, we only consider amorphous carbon because carbonates have been tentatively detected in only one disk to date \citep{Swan+2024}.}
We therefore fix $z = 0.03$ in the main analysis for G29-38. For GD56, where both O and C are not detected, we set $z = 0$.
We also discuss the potential contribution of other moderate conductive materials, such as Fe$_3$O$_4$, to the near-infrared emission in Section~\ref{subsec:other-conductor}.

\subsection{Opacity} 
\label{subsec:opacity-model}
The dust grains are approximated as compact spheres, and their frequency-dependent absorption opacity per dust mass, $\kappa_{\rm abs,\nu}$, is calculated using the publicly available Mie calculation code OpTool \citep{Dominik+2021}.
The particle size distribution is assumed to follow a power law, $n(a) \propto a^{-\alpha}$, between the limits $a_{\rm min} < a < a_{\rm max}$, where $a$ is the dust grain radius. The mass absorption opacity averaged over the grain size distribution is given by
\begin{align}
\overline{\kappa_{\rm abs,\nu}} = \frac{\int_{a_{\rm min}}^{a_{\rm max}} \kappa_{\rm abs,\nu}(a)n(a)da}{\int_{a_{\rm min}}^{a_{\rm max}}n(a)da},
\end{align}
where $\kappa_{\rm abs,\nu}(a)$ is the 
opacity for single-sized particles.
The complex refractive indices ($m = n + ik$) of all amorphous silicates (excluding Mg$_2$SiO$_4$), Mg$_2$SiO$_4$, Fe, \rev{and C} are taken from \citet{Dorschner+1995}, \citet{Jager+2003}, 
\rev{\citet{Ordal+1988}, and the ACAR dataset of \citet{Zubko+1996}}, 
respectively. 

\reveng{
We note that the optical constants of iron-bearing materials can generally depend on the oxidation state of iron.
Silicate minerals consist of [SiO$_4]^{4-}$ tetrahedra as their basic structural units, where the non-bridging oxygens are charge-balanced by metal cations such as Mg$^{2+}$ and Fe$^{2+}$.
While Mg is exclusively present as Mg$^{2+}$, iron can occur as either Fe$^{2+}$ or Fe$^{3+}$ depending on the oxidation state (e.g., in FeO and Fe$_2$O$_3$, respectively).
Because the Fe oxidation state is highly sensitive to the redox conditions during mineral formation, Fe$^{3+}$ can be incorporated under more oxidizing conditions.
\citet{Dorschner+1995} reported that their ferro-magnesium silicate samples 
contain a substantial fraction of Fe$^{3+}$. 
Such Fe$^{3+}$ contamination can alter the near-infrared opacity as well as the shape and peak position of the silicate feature, compared to samples in which all iron is present as Fe$^{2+}$ \citep{Speck+2011}.
In this study, we use the \citet{Dorschner+1995} data for four ferro-magnesium silicate samples:  
MgFeSiO$_4$, Mg$_{0.8}$Fe$_{1.2}$SiO$_4$, Mg$_{0.7}$Fe$_{0.3}$SiO$_3$, 
and Mg$_{0.5}$Fe$_{0.5}$SiO$_3$. 
The potential impact of Fe$^{3+}$ in the \citet{Dorschner+1995} samples on our results is discussed in Section~\ref{subsec:fe-rich-silicate}.
}

\begin{figure}
\centering
\includegraphics[bb=0 0 360 252, width=1.0\columnwidth]{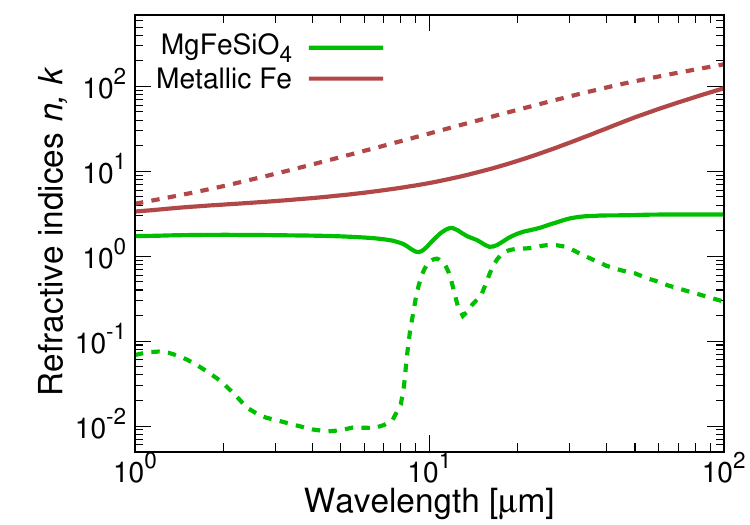}
\caption{Complex refractive indices $m=n+ik$ of MgFeSiO$_4$ (green) and metallic Fe (red). The solid and dotted lines show the real and imaginary parts, $n$ and $k$, respectively.}
\label{fig:refractive}
\end{figure}

The imaginary part $k$ of the complex refractive index represents the material's absorption efficiency. Metallic iron is distinguished from insulating silicates by its significantly high $k$ values at near-infrared wavelengths. 
We illustrate this in Figure~\ref{fig:refractive}, where we plot the refractive indices of silicate MgFeSiO$_4$ and metallic iron as a function of wavelength, $\lambda$.
As the Lorentz model for insulating materials predicts, silicates have $n\simeq O(1)$ and $k \ll 1$ except at feature positions. 
This is because silicates can strongly interact with electromagnetic waves only around specific wavelengths 
\rev{corresponding to lattice vibrational modes, including the Si--O stretching mode at 10~$\mu$m, O--Si--O bending mode at 18~$\mu$m for SiO$_2$, and Mg--O and Fe--O stretching and bending modes for ferro-magnesium silicates. 
In addition to these lattice vibrations, electronic transitions within Fe$^{2+}$ ions can also contribute to the near-infrared absorption of ferro-magnesium silicates, as discussed in Section~\ref{subsec:fe-rich-silicate}.}

In contrast,
 iron has $n \sim k \propto \lambda^{1/2}$ as the Drude model predicts for conducting materials (e.g., \citealt{MiyakeNakagawa1993}).
The $k$ values of iron at $3$--$8\ \mu $m are  three orders of magnitude larger than those of silicates\footnote{\rev{As mentioned above and further discussed in Section \ref{subsec:fe-rich-silicate}, we note the optical constants of Fe-bearing silicates from \citet{Dorschner+1995} included both Fe$^{2+}$ and Fe$^{3+}$, and the difference in absorption relative to metallic iron might be larger those for purely Fe$^{2+}$-bearing silicates.}}. 
This is because free electrons in metal absorb electromagnetic waves across a broader range of wavelengths.

We calculate the effective refractive index and opacity 
of multi-component grains
using the Bruggeman effective medium theory (e.g., \citealt{Bohren&Huffman83}), which does not distinguish between a matrix and inclusions and hence would be applicable for a wide range of metal-to-silicate molar mixing ratios $y$ in Equations (\ref{eq:y-olivine}) and (\ref{eq:y-pyroxene}). 
The mixing rule assumes that the components are physically mixed on a scale smaller than the wavelength under consideration and randomly inhomogeneous.

\begin{figure}
\centering
\includegraphics[bb=0 0 360 252,  width=1.0\columnwidth]{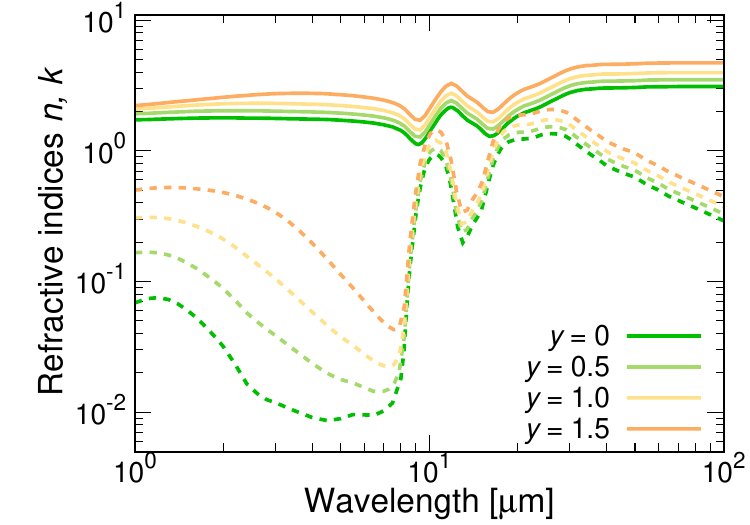}
\caption{Effective refractive indices of materials composed of MgFeSiO$_4$ and metallic iron with various number mixing ratio, $y$. The solid and dotted lines represent $n$ and $k$, respectively. 
}
\label{fig:eff-refractive}
\end{figure}

\begin{figure}
\centering
    \includegraphics[bb=0 0 360 252, width=1.0\columnwidth]{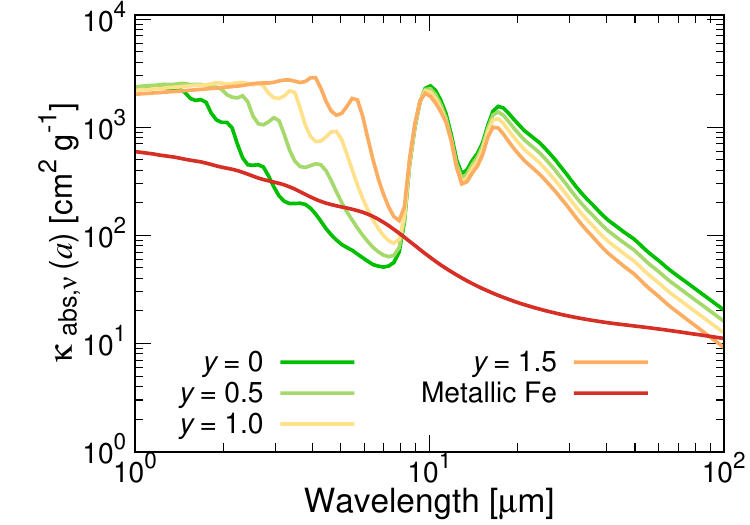}
\caption{Mass absorption opacity of 1 $\mu$m-sized dust grains composed of
MgFeSiO$_4$ and metallic iron with various number mixing ratio, $y$.
}
\label{fig:kappa_mix}
\end{figure}

Figures \ref{fig:eff-refractive} and \ref{fig:kappa_mix} illustrate the effective refractive index and opacity of a mixture of MgFeSiO$_4$ and metallic iron varies with  $y$ in Equation (\ref{eq:y-olivine}). The opacity plotted here assumes $1~\mu$m-sized monodisperse composite grains.
For $y \gtrsim 0.5$, metallic iron significantly enhances $k$ and $\kappa_{\rm abs,\nu}(a)$ in the near-infrared wavelength range.

It is possible that iron and silicates exist in the disks as individual grains.  
However, we find that  the increase in near-infrared dust opacity caused by iron as individual particles is smaller than that caused by iron of the same amount mixed into composite particles (Okuya \& Okuzumi, in prep.).
A noticeable increase in the near-infrared opacity requires $y \gtrsim 3$ in the individual grain case, which is inconsistent with the stellar elemental abundances of G29-38. Therefore, we do not consider such a scenario in our spectral fitting.
\rev{The efficient enhancement of near-infrared opacity in our model would primarily arise from optical coupling, that is, modification of the effective complex refractive index by metallic iron embedded within a silicate medium.
It should be noted that \citet{Speck+2015} discussed the effect of thermal contact between metallic iron and silicate grains in enhancing mid-infrared emission, which is distinct from the focus of this study.}


\subsection{Disk Thermal Emission} \label{subsec:disk-model}
We calculate the thermal emission spectra of the dust disk using the two-layer model developed by \citet{Chiang+1997} and \citet{Chiang+2001}. In this model, the disk is divided into the disk surface, where dust grains are directly irradiated by the central star, and the cooler disk interior, which is heated by the reprocessed starlight from the surface. The disk surface is defined as the layer whose line-of-sight optical depth from the central star reaches unity, i.e., $\tau_{\rm LOS, s, \star} = 1$. The temperature of the surface layer is given by
\begin{equation}
T_{\rm s} = T_{\star} \left[ \frac{1}{4}\left(\frac{R_\star}{r}\right)^2 \left(\frac{
\langle\kappa_{\rm abs} \rangle_{T_{\star}}}{\langle\kappa_{\rm abs} \rangle
_{T_{\rm s}}}\right) \right]^{1/4},
\end{equation}
where $r$ is the disk radius, and $\langle\kappa_{\rm abs} \rangle_T$ is the Planck-mean mass absorption opacity of dust grains at temperature $T$. The temperature of the disk interior is
\begin{align}
T_{\rm i} = T_{\star} \left[\ \frac{1}{2}\left(\frac{R_\star}{r}\right)^2 \sin\theta \right]^{1/4}.
\end{align}
where $\theta$ is 
the mean angle at which the stellar radiation is incident on the disk surface, which can be approximated as \citep{Kusaka+1970,Ruden&Pollack1991}
\begin{align}
\theta &\approx \arctan\left[ \frac{z_{\rm s}}{r} \left(\frac{d\ln z_{\rm s}}{d \ln r}-1 \right) \right] + \arcsin \Bigl(\frac{4R_{\star}}{3\pi r}\Bigr) \notag \\
&\sim \arctan \left(\frac{z_{\rm s}}{r}\right) + \arcsin \Bigl(\frac{4R_{\star}}{3\pi r}\Bigr),
\end{align}
with $\left(\frac{d\ln z_{\rm s}}{d \ln r}-1 \right) \sim O(1)$.
Since the vertical height $z_{\rm s}$ of the $\tau_{\rm LOS, s, \star}=1$ plane in the disks around white dwarfs is uncertain, we simply it to be $z_{\rm s} = r \sin\theta_{\rm hoa}$, where $\theta_{\rm hoa}$ is the half opening angle of the disk surface.

The total flux density $F_\nu$ of the emission from the disk at frequency $\nu$ is the sum of the contributions from the surface and interior, $F_{\nu,\rm s}$ and $F_{\nu,\rm i}$. These can be calculated as
\begin{align}
F_{\nu,\rm s} &= \frac{2\pi\cos i}{d^2}  \int^{r_{\rm out}}_{r_{\rm in}}  B_\nu(T_{\rm s})\ \left(1-e^{-\frac{\tau_{\perp \rm s, \nu}}{\cos i}}\right)
\notag \\
& \quad \times \left(1+e^{-\frac{\tau_{\perp \rm i, \nu}}{\cos i}}\right) \ r dr,
\end{align}
\begin{align}
F_{\nu,\rm i} &= \frac{2\pi\cos i}{d^2}\int^{r_{\rm out}}_{r_{\rm in}}  B_\nu(T_{\rm i})\ \left(1-e^{-\frac{\tau_{\perp \rm i, \nu}}{\cos i}}\right) \ r dr,
\end{align}
respectively, where $B_\nu$ is the Planck function, $i$ is the disk inclination, $r_{\rm in}$ and $r_{\rm out}=r_{\rm in}+\Delta r$ are the inner and outer radii of the disk, and $\tau_{\perp \rm i, \nu}$ and $\tau_{\perp \rm s, \nu}$ are the vertical optical thicknesses of the surface and interior at frequency $\nu$, respectively.
We evaluate $\tau_{\perp \rm s, \nu}$ and  $\tau_{\perp \rm i, \nu}$
as\footnote{
If the surface density of the dust disk is so large that the disk interior layer appears, $z(\tau_{\rm LOS, s, \star}=1)>0$, and, 
\begin{align}
 \tau_{\perp \rm s, \nu} = \frac{\tau_{\perp \rm s, \nu}}{\tau_{\perp \rm s, \star}} \frac{\tau_{\perp \rm s, \star}}{\tau_{\rm LOS, s, \star}} \tau_{\rm LOS, s, \star} \sim \frac{\kappa_{\rm abs, \nu}}{\langle\kappa_{\rm abs} \rangle_{T_{\star}}}  \sin\theta.
\end{align}
Otherwise, all of the vertical region would be treated as the disk surface layer, and 
\begin{align}
 \tau_{\perp \rm s, \nu} = \kappa_{\rm abs, \nu}\frac{\Sigma_{\rm d}{\it (r)}}{2}  .
\end{align}
}
\begin{align}
 \tau_{\perp \rm s, \nu} (r)
=\kappa_{\rm abs, \nu}  \times \min \left\{ \frac{\sin\theta}{\langle\kappa_{\rm abs} \rangle_{T_{\star}}},  \frac{\Sigma_{\rm d}(r)}{2} \right\},
\end{align}
\begin{align}
 \tau_{\perp \rm i, \nu} (r)
=\kappa_{\rm abs, \nu} \Sigma_{\rm d}(r) - 2\tau_{\perp \rm s, \nu},
\end{align}
respectively, where $\Sigma_{\rm d}(r)$ is the surface density of the dust disk.
Since $\sin\theta \sim \sin\theta_{\rm hoa} \ll 1$ (see Table~\ref{tab:model-parameters}) and $\kappa_{\rm abs, \nu}/\langle\kappa_{\rm abs} \rangle_{T_{\star}} \lesssim 1$, the surface is optically thin to its own emission.

We assume that $\Sigma_{\rm d}(r)$ follows a power-law distribution  $ \propto r^{-1}$. Since we consider a radially narrow disk in this model, the effect of the power-law index is not significant (see \citet{Ballering+2022}).
The surface density profile is uniquely determined as a function of $r_{\rm min}$, $r_{\rm out}$, and the total disk mass,
\begin{align}
M_{\rm d} = 2\pi\int^{r_{\rm out}}_{r_{\rm in}}r \Sigma_{\rm d}(r)dr.
\end{align}


We comment on the difference between our disk model and those used in previous studies in Appendix \ref{sec:comparison}.
%

\subsection{Fitting Procedure} \label{subsec:fitting}
\begin{deluxetable}{ccc}
\tablecaption{Model Parameters \label{tab:model-parameters}}
\tablecolumns{5}
\tablewidth{0pt}
\tablehead{\colhead{Parameter} &
\colhead{Description} & Parameter range}
\startdata
$y$ & 
Metal-to-silicate ratio & \begin{tabular}{c}
$0-1.5$ (G29-38) \\ $0-3.5$ (GD56)
\end{tabular}\\
$M_{\rm d}$ (10$^{18}$g) & Dust mass & $3 -3\times 10^{4}$ \\
$r_{\rm in}$ ($R_{\star})$ & Inner radius & $60-110$ \\
$\Delta r$ ($R_{\star}$) & Radial width & $10-40$ \\
$\theta_{\rm hoa}$ (deg.) & Half-opening angle & $1, 4, 16$ \\
$i$ (deg.) & Inclination & $0.5, 30$\\
$a_{\rm min}$ ($\mu$m) & Minimum grain size & $0.3, 1, 3$\\
$a_{\rm max}$ ($\mu$m) & Maximum grain size & $5, 70, 1000$\\
$\alpha$  & Power of grain size distribution & $-2.5, -3.5, -4.5$\\ 
\enddata
\end{deluxetable}

We search for the best-fit model to the observational spectra of GD29-38 and GD56 (Figure \ref{fig:binning}) by varying the metal-to-silicate molar mixing ratio $y$ and 8 physical parameters involved in our disk thermal emission model.
Table \ref{tab:model-parameters} summarizes our fitting parameters and their variation ranges.
Because the pollutant Fe/Si for GD56 is inferred to be larger than that for G29-38 (Table~\ref{tab:ss-elements}), we assume higher values of $y$ for GD56.
The parameter space is sampled using 116640 and 233280 grid points for G29-38 and GD56, respectively.

We evaluate the goodness-of-fit of the spectra by minimizing the reduced $\chi^2$ values defined as
\begin{align}
\chi^2 &= \frac{1}{q}\sum_{\rm i}^{n_{\rm data}} w\frac{(F_{\rm model}(\lambda)- F_{\rm obs}(\lambda))^2}{\sigma_{\rm obs}(\lambda)^2}, \\
& \quad q = \sum_{\rm i}^{n_{\rm data}}  w - n_{\rm d.o.f},
\label{eq:chi}
\end{align}
where $\sigma_{\rm obs}$ is a 1-$\sigma$  observational noise and $w$ is a weighting factor, which we set as follows.
As the metallic iron and silicates influence the absorption opacity in the near- and mid-infrared, respectively (Figure \ref{fig:kappa_mix}), the near-infrared photometric data and mid-infrared binned spectral data are equally important to determine the best-fit metal-to-silicate ratio.
Therefore, we adjust the value of $w$ so that both data sets are equally weighted in the $\chi^2$ evaluation.
For G29-38, we set $w=1$ for all data points.
For GD56, we set $w=1$ for photometric data and $w=2$ for the binned spectral data, as $\sigma_{\rm obs}$ of binned spectra data is larger than that of photometric data by a factor of a few.

\section{Results} \label{sec:results}
As noted in Section~\ref{subsec:opacity-model}, Fe metal significantly contributes to the near-infrared opacity.
In this section, we constrain the best-fit metal-to-silicate molar mixing ratio $y$ (Equation~\ref{eq:y-olivine}), in the dust surrounding G29-38 (Section~\ref{subsec:Fe-G29-38}) and GD56 (Section~\ref{subsec:Fe-GD56}).

\subsection{Metallic Iron Abundance in Dust around G29-38} \label{subsec:Fe-G29-38}

\begin{figure*}
\centering
\includegraphics[bb=0 0 720 540, keepaspectratio, scale=0.55]{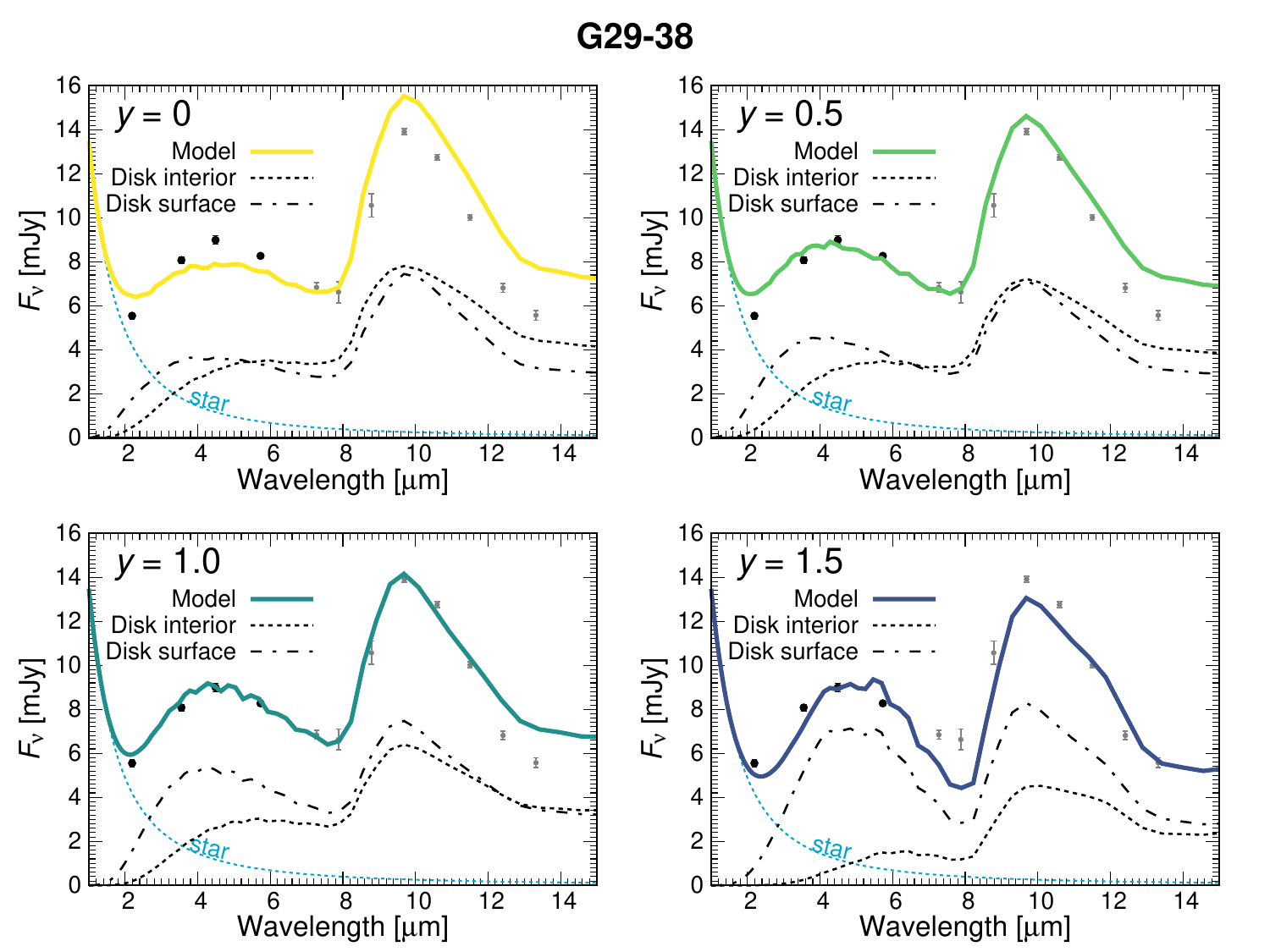}
     \caption{
    Comparison between the model disk spectra (solid color lines) at a given metal-to-silicate molar mixing ratio ($y$) and the observational data (black and gray dots) for G29-38 assuming MgFeSiO$_4$ silicate composition. The model spectra are the sum of the emission from the star (cyan dotted lines), disk interior (black dotted lines), and disk surface (black dotted-dashed).}
     \label{fig:Fe-dep-G29-38}
  \end{figure*}

First, we fix the silicate composition to MgFeSiO$_4$ and examine how the goodness-of-fit of the G29-38 spectrum depends on $y$.
Figure~\ref{fig:Fe-dep-G29-38} shows the best-fit model spectrum for four different, given values of $y$. The fitting parameters other than $y$ are varied freely, and the derived best-fit values are listed in Table \ref{tab:param-MgFeSiO4-G29}.
A larger $y$ produces a larger disk surface emission in the near-infrared, resulting in a better fit to observed excess around $\lambda\sim 5\ \mu$m. 
Additionally, a moderate amount of metallic iron also reproduces the 10 $\mu$m silicate feature well. 
In contrast, iron-free dust has a smaller near-infrared opacity. Adopting a smaller $r_{\rm in}$ and hence a higher $T_{\rm s}$ partially compensates for this, but the resulting flux at $\lambda \sim 5\ \mu$m remains smaller than the observed one. Moreover, this excessively hot disk results in an overprediction of the 10 $\mu$m silicate feature.
Dust with a too large value of $y$ provides a less sharp silicate feature in the opacity (Figure~\ref{fig:kappa_mix}), leading to a worse fit to the mid-infrared spectrum.

\begin{figure}
\centering
\includegraphics[bb=0 0 360 252, width=1.0\columnwidth]{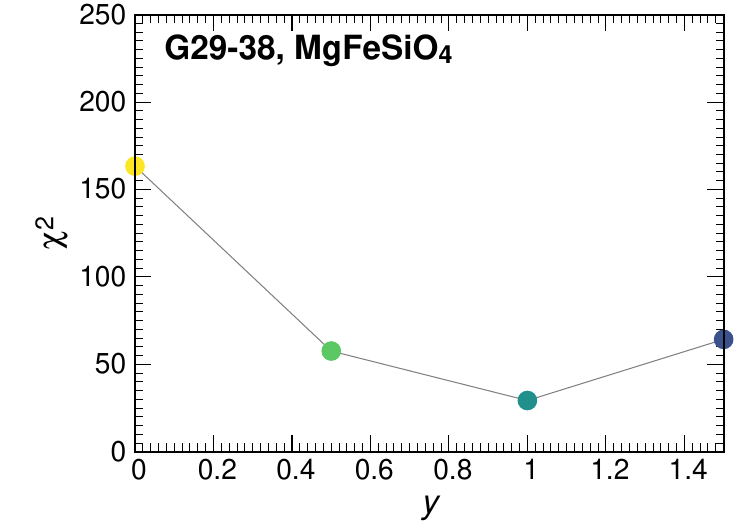}
\caption{$\chi^2$ value of model disk spectra for G29-38 assuming MgFeSiO$_4$ silicate composition as a function of the metal-to-silicate ratio ($y$). 
}
\label{fig:chi_wo_color_G29-38_MgFeSiO4}
\end{figure}

Figure \ref{fig:chi_wo_color_G29-38_MgFeSiO4} shows corresponding $\chi^2$ values calculated from  Equation~(\ref{eq:chi}).
The $\chi^2$ values quantitatively indicate that the spectra of the iron-bearing dust with $y \ge 0.5$ provide better fits to the G29-38 data, with $y\sim 1$ yielding the lowest $\chi^2$ value.

\begin{deluxetable*}{ccccccccc}
\tablecaption{Best-fit Disk Parameters for G29-38 Assuming MgFeSiO4 Dust \label{tab:param-MgFeSiO4-G29}}
\tablecolumns{5}
\tablewidth{0pt}
\tablehead{
\colhead{\begin{tabular}{c}$y$ \\ \end{tabular} }&
\colhead{\begin{tabular}{c} $M_{\rm d}$ \\ ($10^{19}$ g) \end{tabular}}
&
\colhead{\begin{tabular}{c} $r_{\rm in}$ \\ ($R_{\star}$)\end{tabular}} &
\colhead{\begin{tabular}{c} $\Delta r$ \\ ($R_{\star}$)\end{tabular}}&
\colhead{\begin{tabular}{c} $\theta_{\rm hoa}$\\ ($^\circ$) \end{tabular}} &
\colhead{\begin{tabular}{c} $i$ \\ ($^\circ$) \end{tabular}} &
\colhead{\begin{tabular}{c} $a_{\rm min}$ \\ ($\mu$m) \end{tabular}} &
\colhead{\begin{tabular}{c} $a_{\rm max}$ \\ ($\mu$m) \end{tabular}}&
\colhead{\begin{tabular}{c} $\alpha$ \\ \end{tabular}} 
}
\startdata
0  & 10 & 80 & 10 & 16 & 30 & 0.3 & 1000 & -3.5\\
0.5  & 10 & 90 & 10 & 16 & 30 & 0.3 & 1000 & -3.5\\
1.0  & 10 & 110 & 10 & 16 & 0.5 & 0.3 & 1000 & -3.5\\
1.5  & 1.0 & 110 & 30 & 4 & 0.5 & 1.0 & 5 & -2.5\\
\enddata
\end{deluxetable*}

\begin{figure*}[t]
\centering
     \includegraphics[bb=0 0 864 518, keepaspectratio, scale=0.55]{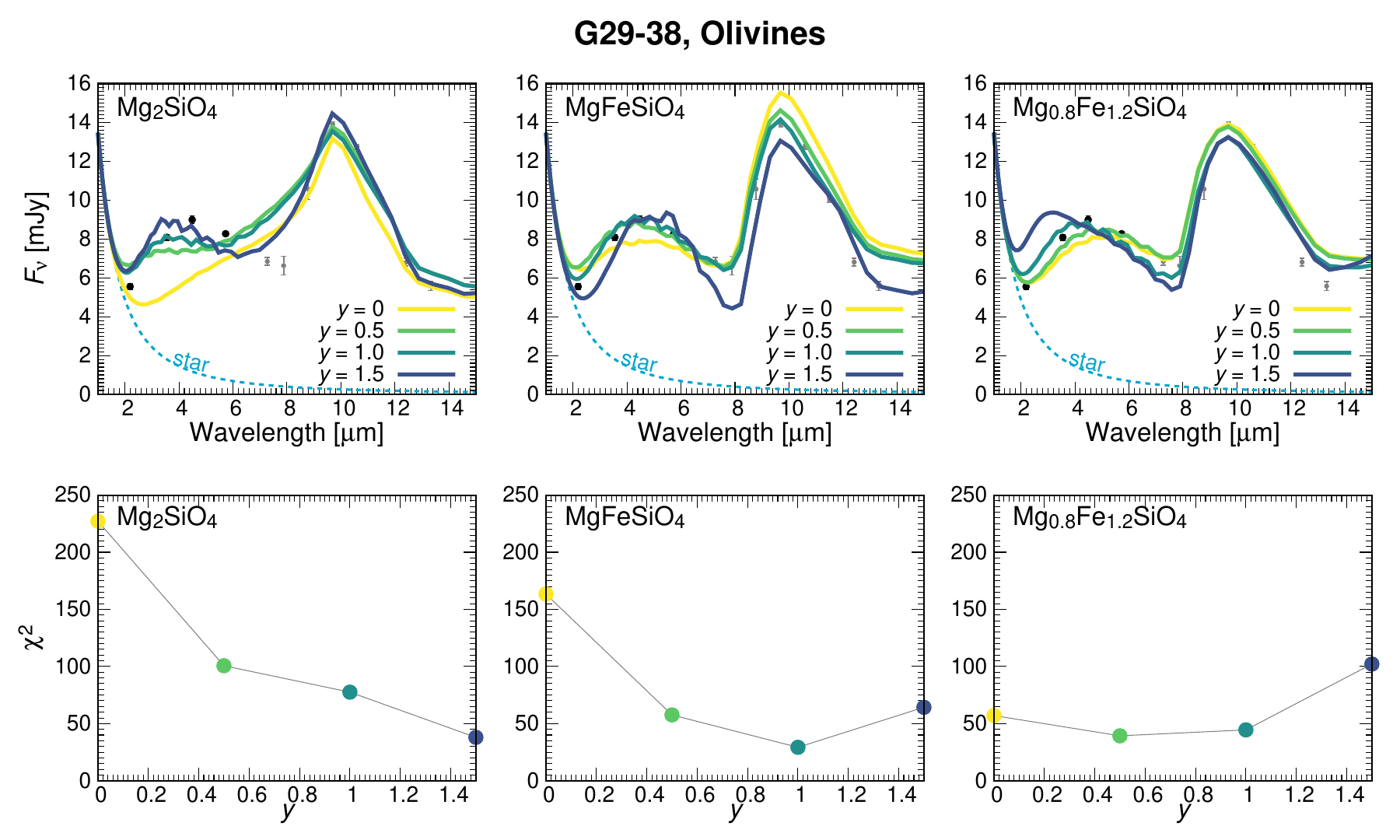}
 \\
     \includegraphics[bb=0 0 864 518, keepaspectratio, scale=0.55]{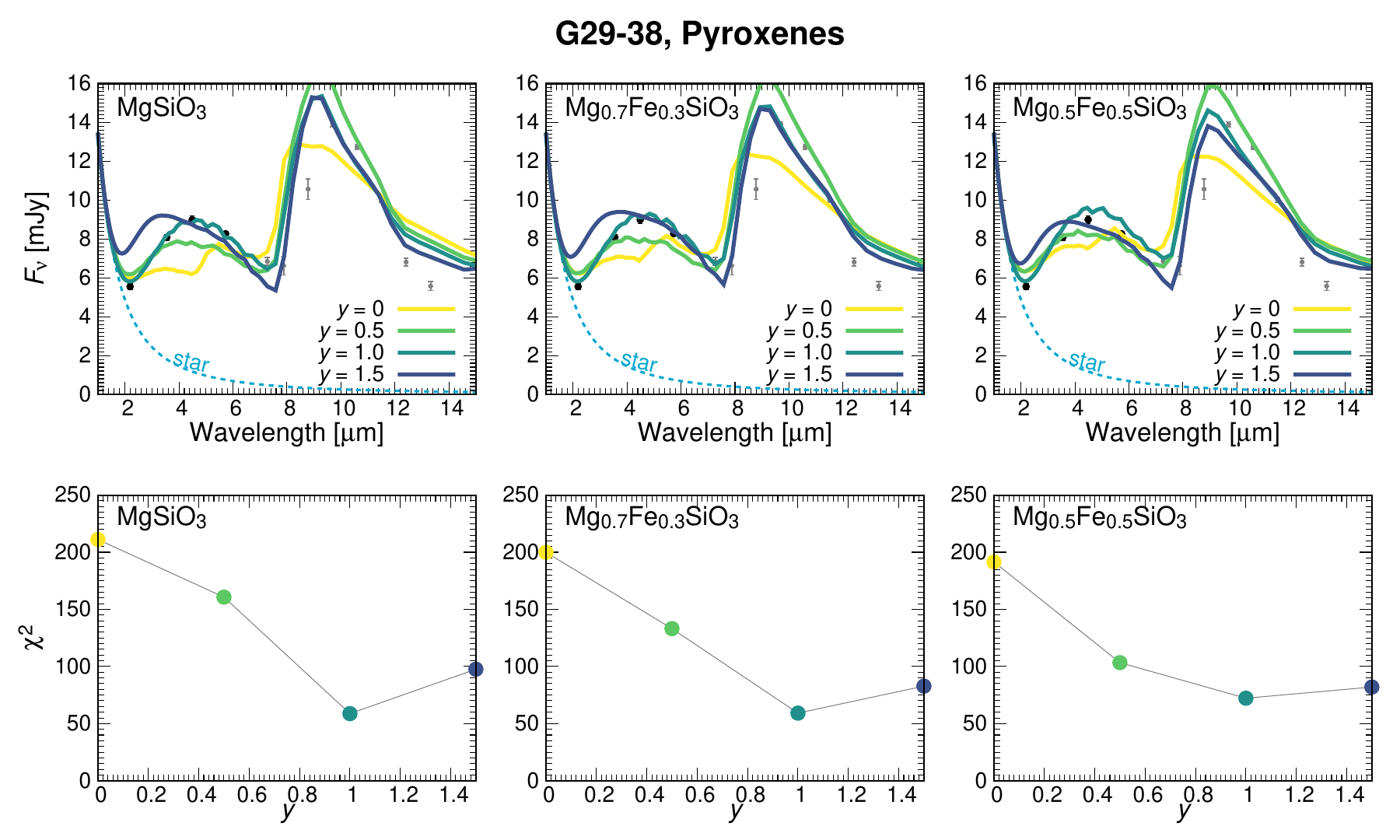}
\caption{
Best-fit G29-38 spectra and their $\chi^2$ values from models with different silicate compositions and metal-to-silicate ratios $y$. The Fe/Mg ratio of the silicates, $x$, increases from the left to  right panels. 
The cyan dotted lines and the dots with errorbars show model spectra of the star and the observational data, respectively.
}
\label{fig:G29-38-silicate}
\end{figure*}
Next, we examine whether the assumed silicate composition affects our conclusion that iron-containing dust can better reproduce the observed disk spectrum around G29-38.
We perform similar analyses for amorphous olivines and pyroxenes with various Mg/Si ratios $x$ (Equation~\ref{eq:y-olivine} and Equation~\ref{eq:y-pyroxene}).
Figure~\ref{fig:G29-38-silicate} shows the best-fit spectra and their $\chi^2$ values for different silicate compositions and $y$ values.
For all silicate compositions excluding Mg$_{0.8}$Fe$_{1.2}$SiO$_4$, 
\rev{metallic-}iron-bearing dust better reproduces the observed spectrum.
The dust with Mg$_{0.8}$Fe$_{1.2}$SiO$_4$ reproduces the observed spectrum equally well with and without metallic iron.
Olivine silicates better match the silicate feature than pyroxenes, consistent with previous work (e.g., \citealt{Jura+2009AJ}). The difference in the silicate feature shape at $\lambda\sim$ 6--8 $\mu$m for Mg$_2$SiO$_4$ compared to other olivines reflects its imaginary part of refractive index, which exhibits a monotonical decrease in that wavelength range \citep{Jager+2003}.

\subsection{Metallic Iron Abundance in Dust around GD56} \label{subsec:Fe-GD56}
We analyze the GD56 spectrum with the same spectral fitting procedures as in Section~\ref{subsec:Fe-G29-38}. 
As the inferred pollutant Fe/Si ratio for GD56 is substantially higher than that for G29-38 (Table \ref{tab:ss-elements}), we extend the $y$ range in our spectral fitting up to 3.5.

\begin{figure*}[t]
    \centering 
    \includegraphics[bb = 0 0 1008 288, keepaspectratio, scale=0.5]{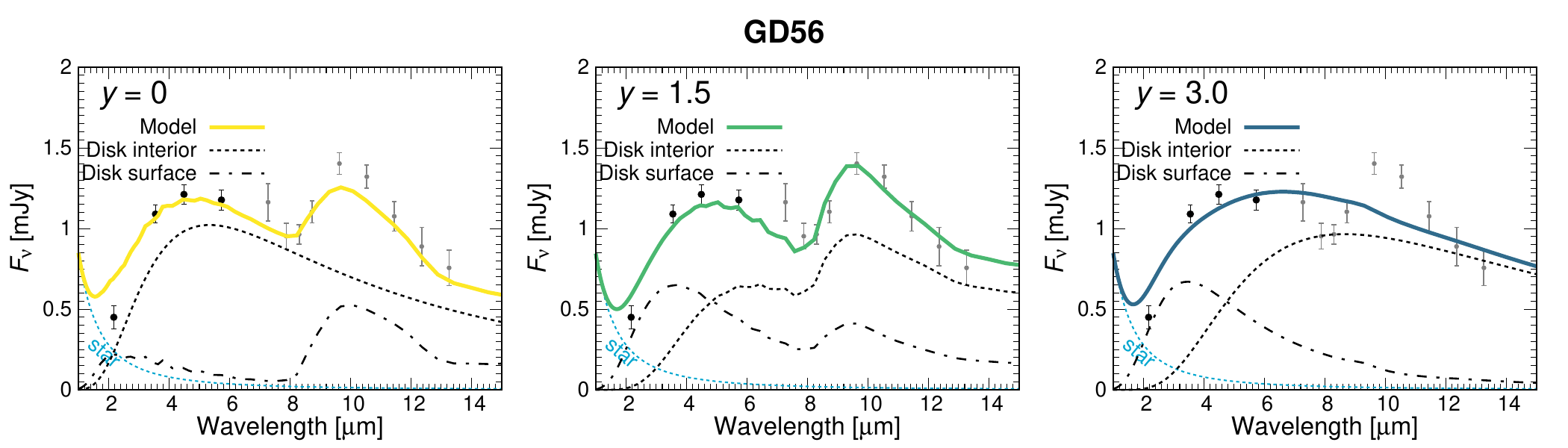}
     \caption{
    Same as Figure~\ref{fig:Fe-dep-G29-38} but for GD56.}
     \label{fig:Fe-dep-GD56}
  \end{figure*}

\begin{deluxetable*}{ccccccccc}
\tablecaption{Best-fit Disk parameters for GD56 Assuming MgFeSiO4 \label{tab:param-MgFeSiO4-GD56}}
\tablecolumns{1}
\tablehead{
\colhead{\begin{tabular}{c} $y$ \\ \end{tabular} }&
\colhead{\begin{tabular}{c} $M_{\rm d}$ \\ ($10^{19}$ g) \end{tabular}}
&
\colhead{\begin{tabular}{c} $r_{\rm in}$ \\ ($R_{\star}$)\end{tabular}} &
\colhead{\begin{tabular}{c} $\Delta r$ \\ ($R_{\star}$)\end{tabular}}&
\colhead{\begin{tabular}{c} $\theta_{\rm hoa}$\\ ($^\circ$) \end{tabular}} &
\colhead{\begin{tabular}{c} $i$ \\ ($^\circ$) \end{tabular}} &
\colhead{\begin{tabular}{c} $a_{\rm min}$ \\ ($\mu$m) \end{tabular}} &
\colhead{\begin{tabular}{c} $a_{\rm max}$ \\ ($\mu$m) \end{tabular}}&
\colhead{\begin{tabular}{c} $\alpha$ 
\end{tabular}} 
}
\startdata
0  & 320 & 80 & 10 & 16 & 0.5 & 1.0 & 5.0 & -2.5 
\\
1.5  & 32 & 90 & 40 & 4 & 30 & 0.3 & 1000 & -3.5 
\\
3.0  & 32 & 100 & 30 & 4 & 0.5 & 0.3 & 71 & -4.5 
\enddata
\end{deluxetable*}

\begin{figure}
\centering
    \includegraphics[bb=0 0 360 252, width=1.0\columnwidth]{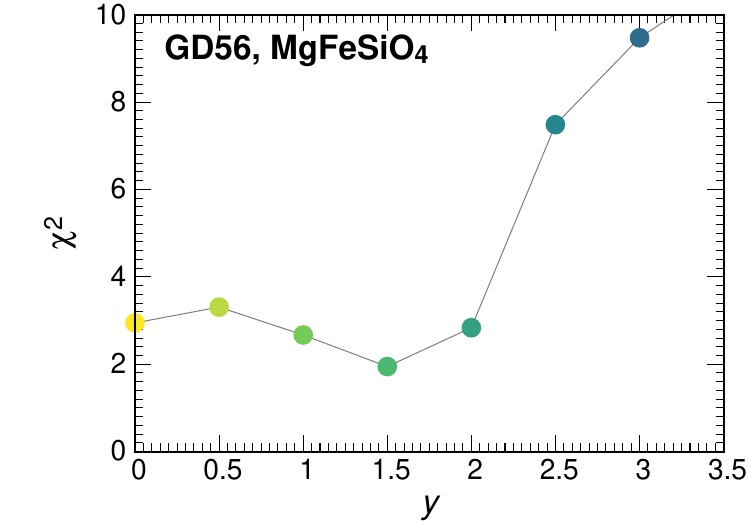}
\caption{Same as Figure~\ref{fig:chi_wo_color_G29-38_MgFeSiO4} but for GD56.
}
\label{fig:chi_wo_color_GD56_MgFeSiO4}
\end{figure}

We again begin by fixing the silicate composition to MgFeSiO$_4$ and to study how the best-fit spectrum for GD56 depends on $y$. 
The best-fit spectra for $y = 0, 1.5$, and $3$ are shown in Figure ~\ref{fig:Fe-dep-GD56}, and the best-fit disk physical parameters for each $y$ are listed in Table \ref{tab:param-MgFeSiO4-GD56}.
The $\chi^2$ values versus $y$ are plotted in Figure \ref{fig:chi_wo_color_GD56_MgFeSiO4}.

The $\chi^2$ values indicate that the model with $y\sim 1.5$ best reproduces the observed GD56 spectrum (Figure~\ref{fig:chi_wo_color_GD56_MgFeSiO4}).
Increasing the abundance of metallic iron enhances the near-infrared opacity (Fig~\ref{fig:kappa_mix}), resulting in a stronger near-infrared emission from the disk surface (Figure~\ref{fig:Fe-dep-GD56}). 
Additionally, a moderate amount of metallic iron well reproduces the 10 $\mu$m silicate feature.
On the other hand, the model with $y = 0$ underpredicts the 10 $\mu$m silicate feature, leading to a larger $\chi^2$ value than with $y = 1.5$ (Figure \ref{fig:chi_wo_color_GD56_MgFeSiO4}). 
The model with too much metallic iron, $y = 3$, also results in a larger $\chi^2$ because the opacity of the iron-dominating dust provides a more featureless spectrum. 

These results are similar to those for G29-38.
However, the best-fit disk physical parameters are different between G29-38 and GD56.
For instance, the best-fit $M_{\rm d}$ value for GD56 is slightly larger than that for G29-38.
Additionally, the best-fit $r_{\rm in}/R_{\star}$ for GD56 is larger than that for G29-38. This results in the higher best-fit disk temperatures for GD56. 

We caution that all the $\chi^2$ values for $0 \leq y \leq 2$ fall below 4 and show little dependence on $y$.
Indeed, the models with $0 \leq y \leq 2$ reproduce the observed spectrum equally well within the 1-$\sigma$ observational uncertainty.
Thus, higher-precision infrared spectroscopy is needed to place a stronger constraint on the best-fit $y$ value for GD56. 
This may be achievable with the upcoming observational data at $\lambda = 5-28~\mu$m using the JWST MIRI \citep{Kate+2023jwst}. Additionally, future observations with the JWST Near InfraRed Spectrograph (NIRSpec), which covers $0.6-5.3~\mu$m, could further refine these constraints.

\begin{figure*}[t]
\centering
     \includegraphics[bb=0 0 864 518, keepaspectratio, scale=0.55]{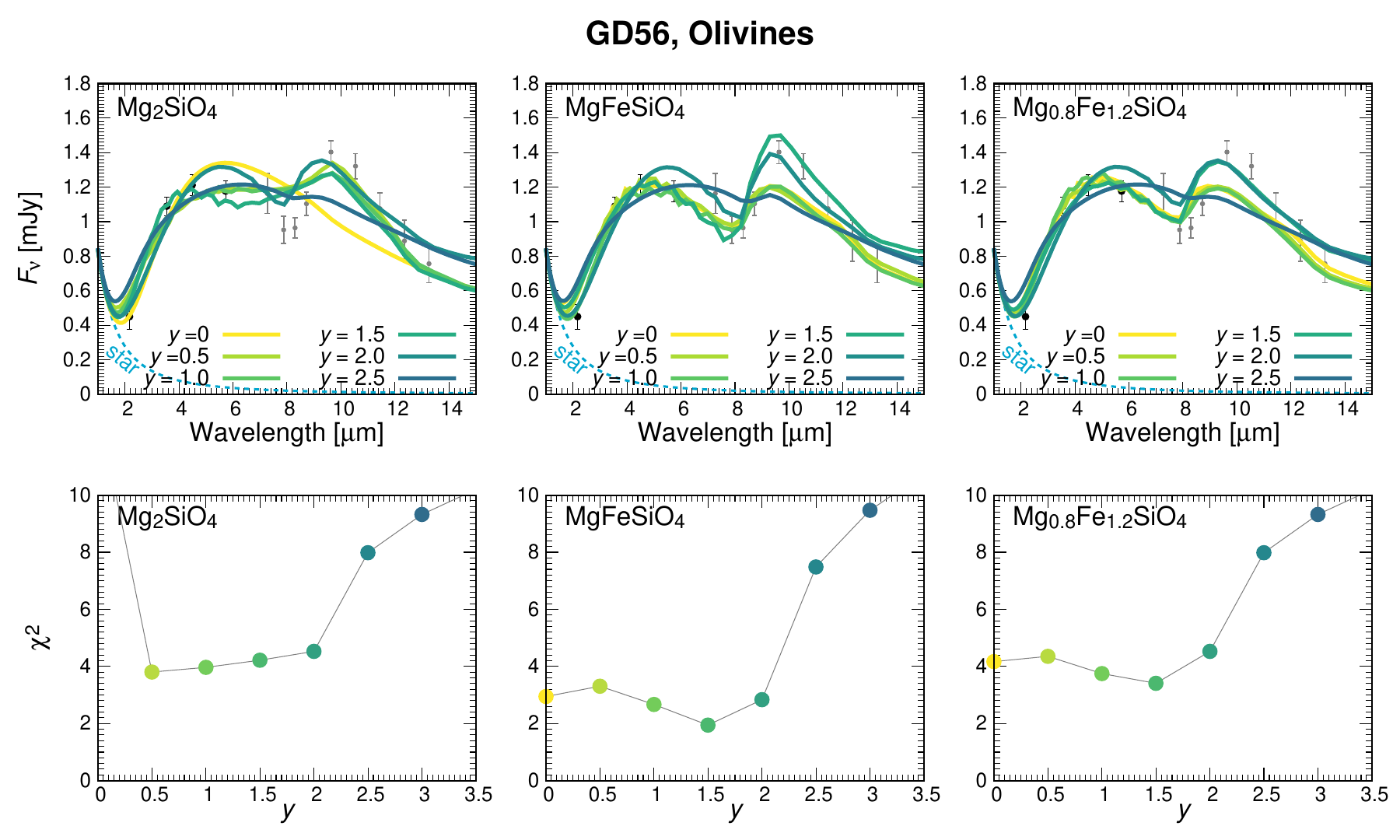}
 \\
     \includegraphics[bb=0 0 864 518, keepaspectratio, scale=0.55]{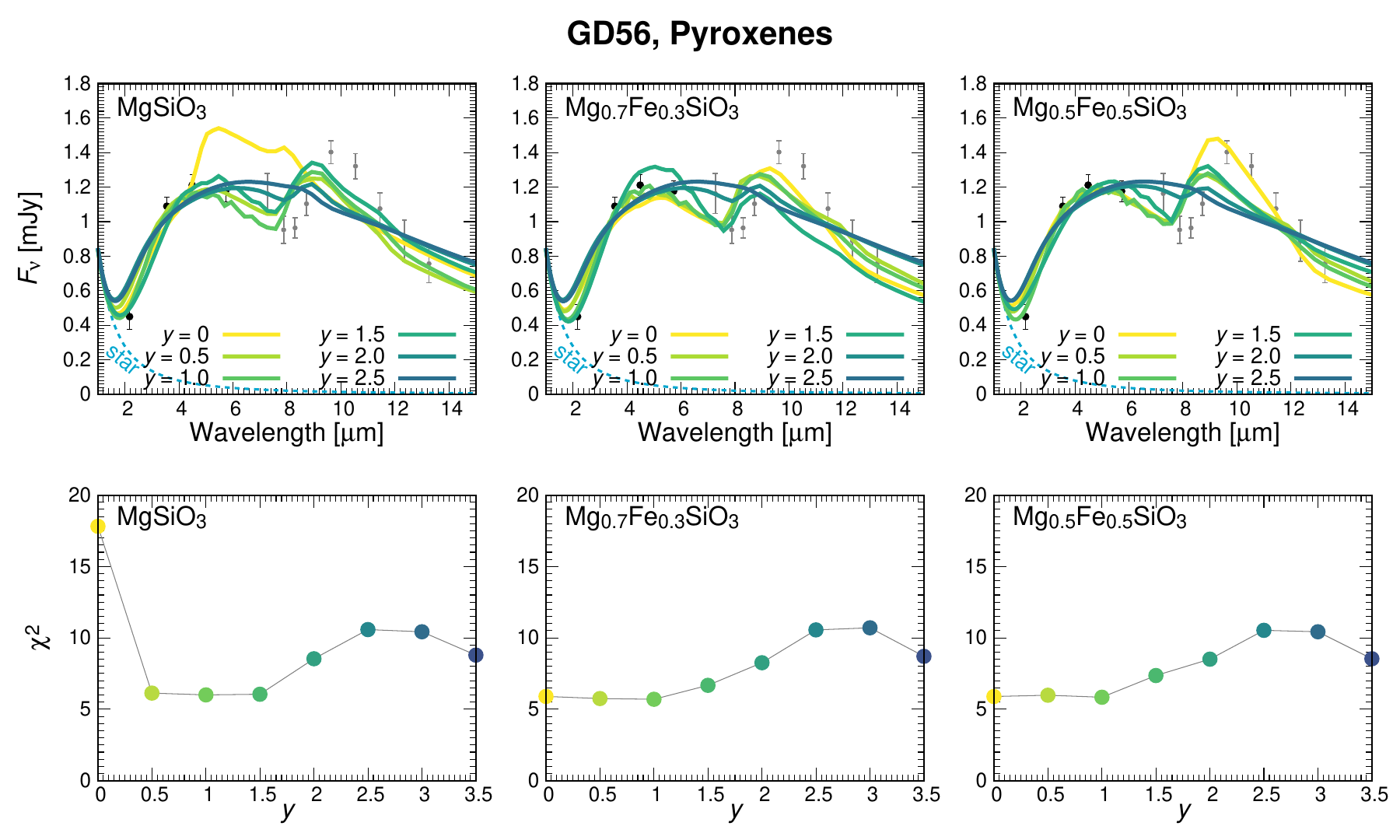}
\caption{Same as Figure \ref{fig:G29-38-silicate} but for GD56.
}
\label{fig:GD56-silicate}
\end{figure*}

We also examine the dependence of the best-fit $y$ value on the assumed silicate composition.
Figure \ref{fig:GD56-silicate} shows the best-fit spectra and their $\chi^2$ values for different $y$ values and silicate compositions.
All silicate compositions excluding non-Mg-pure pyroxenes (i.e., Mg$_{0.7}$Fe$_{0.3}$SiO$_3$ and Mg$_{0.5}$Fe$_{0.5}$SiO$_3$) prefer an addition of metallic iron by a factor of 1.2 to 4 in the $\chi^2$ values.
In particular, Mg-pure silicates (i.e., Mg$_2$SiO$_4$ and MgSiO$_3$) strongly favor $y\gtrsim 0.5$.
However, excluding the Mg-pure silicate case, the models with $0 \leq y \lesssim 2$ reproduce the observed spectrum equally well within the 1-$\sigma$ observational uncertainty.
%
A re-analysis of the upcoming JWST MIRI data will give more stringent constraints on the silicate composition.

\section{Discussion} \label{sec:discussion}

\subsection{Correlation between the Iron Abundance in Dust and in Stellar Atmospheres} \label{subsec:correlation}

The spectral fitting presented in Section \ref{sec:results} have provided the best-fit metal-to-silicate mixing ratio ($y$) of circumstellar dust around G29-38 and GD56.
In this section, we compare these values with the 
pollutant iron abundances calculated from the stellar atmospheres taking into account $t_{\rm sink}$ (Table \ref{tab:ss-elements}). 
By construction, our spectral model with the nominal dust composition assumes the relative abundances of Mg/Si, O/Si, and C/Si consistent with those derived from the stellar atmospheres (see Section \ref{subsec:dustcomp}).

\begin{figure*}[t]
\centering
     \includegraphics[bb= 0 0 864 518, keepaspectratio, scale=0.55]{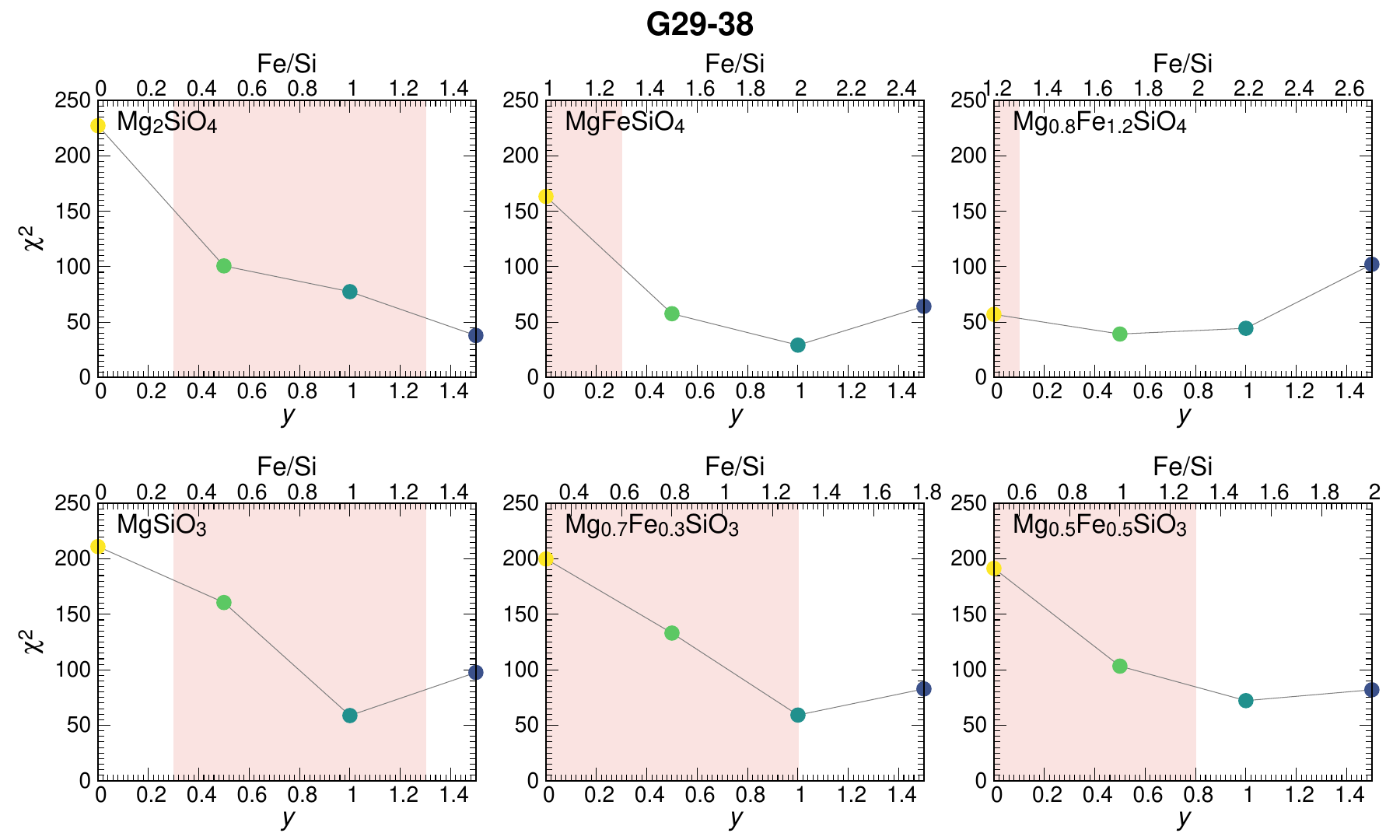}
\caption{
$\chi^2$ values of the model spectra for G29-38 for different silicate compositions and metal-to-silicate ratios $y$ shown in Figure \ref{fig:G29-38-silicate}, but here over-plotting the 1$\sigma$ ranges of the Fe/Si ratio calculated from the stellar atmospheres considering $t_{\rm sink}$ (pink shaded regions). 
The metal-to-silicate mixing ratio ($y$) on the lower $x$-axis corresponds to the Fe/Si value on the upper $x$-axis.
}
\label{fig:G29-38_vs_stellar}
\end{figure*}

\begin{figure*}[t]
\centering
     \includegraphics[bb= 0 0 864 518, keepaspectratio, scale=0.55]{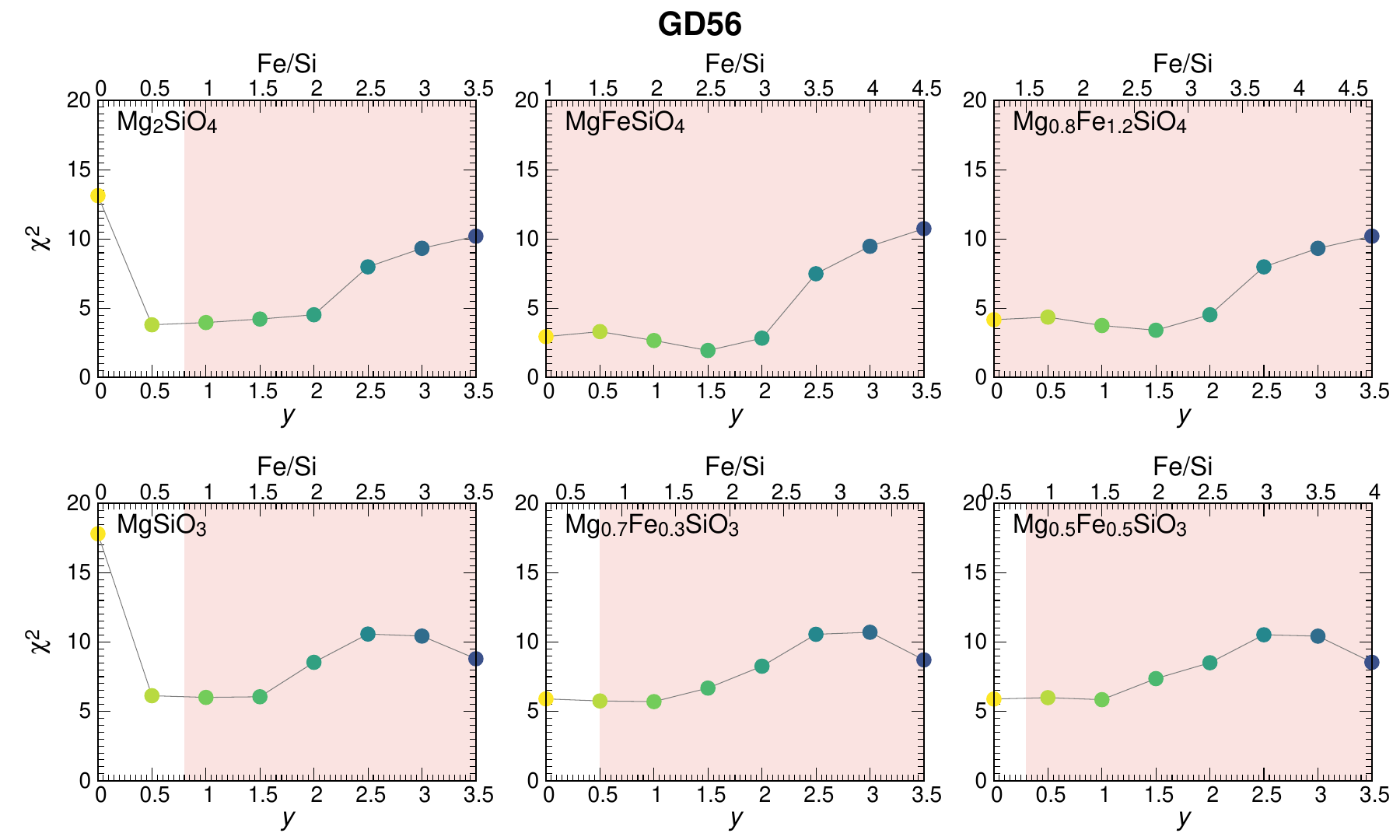}
\caption{Same as Figure \ref{fig:G29-38_vs_stellar} but for GD56.
}
\label{fig:GD56_vs_stellar}
\end{figure*}

Here, we compare the Fe/Si values of the circumstellar dust and stellar atmospheres (considering $t_{\rm sink}$ for the stellar atmospheres).
The Fe/Si ratios of the circumstellar dust consisting of olivines and pyroxenes in addition to metallic iron are given by,
\begin{align}
{\rm (Fe/Si)_{dust}} &= (2 - x) + y, \\
{\rm (Fe/Si)_{dust}} &= (1 - x) + y,
\end{align}
respectively (see Equations (\ref{eq:y-olivine}) and (\ref{eq:y-pyroxene})).
Figure \ref{fig:G29-38_vs_stellar} and \ref{fig:GD56_vs_stellar} compare the best-fit $y$ and the corresponding (Fe/Si)$_{\rm dust}$ values with the inferred ranges of the stellar (Fe/Si)$_{\star}$ values, represented by the shaded areas, for G29-38 and GD56, respectively.
The iron abundance in the circumstellar dust can be considered consistent with that in the stellar atmospheres if the $y$ value corresponding to the lowest $\chi^2$ lies within the shaded range.
In this case, the gas flow accreting onto the stellar surface would have a composition similar to the dust in the disk around the star.

For G29-38, three dust compositions MgSiO$_3$+Fe, Mg$_{0.7}$Fe$_{0.3}$SiO$_3$+Fe, and Mg$_{0.8}$Fe$_{1.2}$SiO$_4$ with no metallic iron yield the best-fit (Fe/Si)$_{\rm dust}$ values consistent with (Fe/Si)$_{\star}$. 
The other silicate compositions predict $\rm (Fe/Si)_{dust} >(Fe/Si)_{\star}$, indicating either that they do not represent the actual dust composition in the G29-38 disk, or that some mechanism within the disk depletes the accreting flow in Fe (or enriches it in Si) relative to the disk dust.
For instance, metallic iron and silicate may have different accretion rates if the angular momentum exchange between gas and dust is efficient.
\citet{Okuya+2023} simulated the coupled evolution of water vapor and silicate dust/gas and has shown that  the water vapor accelerates the accretion of the silicate dust through gas drag. In contrast, the accretion of the water vapor itself is suppressed due to the back-reaction from the silicate dust, leading to an increased silicate-to-water vapor ratio compared to the initial disk composition. 
Since the sublimation temperature of metallic iron is approximately 300 K lower than those of silicates \citep[e.g.,][]{Rafikov+2012}, metallic iron sublimates while silicates remain in solid form. Therefore, metallic iron might play a role similar to water vapor, potentially decreasing the iron-to-silicate ratio in the accretion flow. Theoretical calculations to confirm this possibility are left for future work. 
%

For GD56, the iron abundance in the dust is consistent with that in the stellar atmosphere (Figure \ref{fig:GD56_vs_stellar}).
In particular, Mg-pure silicates strongly favor dust with metallic iron ($\rm (Fe/Si)_{dust} \gtrsim 0.5$), which is consistent with $\rm (Fe/Si)_{\star} \ge 0.8$. 
\rev{Fe-rich olivines without metallic iron are also compatible within the 1-$\sigma$ uncertainty range. }
This consistency indicates that neither the dust composition nor the stellar atmospheric composition supports a value of $\rm Fe/Si \sim 0$.
However, making a more precise comparison between the iron abundances in the disk dust and stellar atmosphere is difficult with the existing stellar atmospheric data because of the large observational uncertainties of the stellar elemental abundances.

As the observational uncertainties are large for individual systems, it is crucial to study a large number of systems to gain a statistical understanding of the correlation between the Fe/Si values of the circumstellar dust and stellar atmospheres. \rev{Recent and upcoming survey results from JWST \citep[e.g.,][]{Farihi+2025}} will greatly expand the sample of polluted white dwarfs with observed infrared emission spectra from their disks. Furthermore, ongoing spectroscopic surveys using SDSS and DESI, and WEAVE will increase the sample of white dwarfs with measurements of multiple elements' abundance in their atmospheres.



\subsection{Fe-rich silicate as an alternative solution: the role of Fe$^{2+}$ and Fe$^{3+}$ in silicate opacity}
\label{subsec:fe-rich-silicate}

\reveng{
As described in Section~\ref{subsec:opacity-model}, we adopt the optical constants measured by \citet{Dorschner+1995} for MgFeSiO$_4$, Mg$_{0.8}$Fe$_{1.2}$SiO$_4$, Mg$_{0.7}$Fe$_{0.3}$SiO$3$, and Mg$_{0.5}$Fe$_{0.5}$SiO$_3$.
These samples contained a substantial fraction of Fe$^{3+}$, with an approximate Fe$^{2+}$:Fe$^{3+}$ ratio of $\sim 1:2$.
We here discuss how the use of the data from Fe$^{3+}$-bearing samples may influence the results of our spectral fitting.
}

\reveng{
In general, the absorption of light in a solid arises from several mechanisms.
Vibrational transitions of molecular or atomic lattices produce infrared absorption in insulating materials such as silicates, while free-electron transitions in conductive materials such as metallic iron generate broad absorption across the visible to infrared wavelength range (see Section~\ref{subsec:opacity-model}).
In addition to these processes, electronic transitions between partially filled orbitals within the same atom can also contribute to absorption.
According to crystal field theory, when Fe is incorporated into the coordination environment surrounded by O$^{2-}$ anions in silicates, the energy levels of its $d$ orbitals split \citep[e.g.,][]{Burns1993}. 
While Fe$^{3+}$ has a half-filled, relatively stable electronic configuration, Fe$^{2+}$ more readily undergoes electronic transitions whose energies correspond to near-infrared wavelengths ($\sim 1~\mu$m). 
As a result, Fe$^{2+}$-bearing silicates exhibit enhanced near-infrared opacity \citep{Zeidler+2011}. 
This absorption mechanism is characteristic of Fe$^{2+}$ and is largely absent in Fe$^{3+}$-dominated phases such as Fe$_2$O$_3$ \citep{Morris+1985}.
}

\reveng{
Therefore, if optical constants in which all iron is present as Fe$^{2+}$ were used, the near-infrared emission from Fe-bearing silicates would likely be even stronger. 
As a result, the best-fit metal-to-silicate mixing ratios ($y$ value) for the Fe-bearing silicates 
(MgFeSiO$_4$, Mg$_{0.8}$Fe$_{1.2}$SiO$_4$, Mg$_{0.7}$Fe$_{0.3}$SiO$_3$, and Mg$_{0.5}$Fe$_{0.5}$SiO$_3$) could become smaller than derived in Section~\ref{sec:results}.
If this is the case, the correlation between the iron abundance in dust and in stellar atmosphere discussed in Section~\ref{subsec:correlation} would be modified.  
In the particular case of G29-38, where the data quality is higher, several of these Fe-bearing silicates yield 
$\mathrm{(Fe/Si)_{dust}} > \mathrm{(Fe/Si)_\star}$. 
Using Fe$^{2+}$-only optical data could move 
$\mathrm{(Fe/Si)_{dust}}$ closer to $\mathrm{(Fe/Si)_\star}$.
Thus, Fe-rich silicate compositions without metallic iron cannot be ruled out based on consistency with the stellar atmospheric abundances.
We note that an interesting example already appears in our current analysis:
the Mg$_{0.8}$Fe$_{1.2}$SiO$_4$ optical constants reproduce the observed spectrum
of G29$-$38 well even when almost no metallic iron is included ($y \sim 0$),
and its inferred $\mathrm{(Fe/Si)_{dust}}$ is already close to $\mathrm{(Fe/Si)_\star}$.
}

\reveng{
In addition, the presence of Fe$^{2+}$ and Fe$^{3+}$ can also affect the shape and peak position of the silicate feature. 
This is because Fe$^{2+}$ acts as a network modifier}\footnote{\reveng{A network modifier increases the number of non-bridging oxygens that are not part of the SiO$_4$ tetrahedral framework. 
In contrast, a network former reduces non-bridging oxygens. 
The ratio of non-bridging oxygens to metal cations serves as an indicator of the degree of polymerization: 
a larger value corresponds to a more open, glassy, and softer structure, whereas a smaller value corresponds to a more polymerized and rigid structure.}}
\reveng{, whereas Fe$^{3+}$ behaves as a network former, thereby modifying the degree of polymerization \citep{Speck+2011}. 
These structural effects can influence the goodness of the silicate-feature fitting.
Therefore, assessing whether the best-fit $y$ value would actually become smaller when all iron is present as Fe$^{2+}$ requires an analysis using optical constants measured from glass samples synthesized under fully controlled redox conditions.
}

\reveng{
If silicates with all iron present as Fe$^{2+}$ can also reproduce the observed spectra, 
the above results provide an additional perspective on the conventional understanding of the so-called dirty silicates.
To explain the insufficient near-infrared opacity of pure magnesium silicates, models invoking dirty silicates, heterogeneous grains in which metallic iron or other absorbing materials are mixed into the silicate, have been widely adopted \citep[e.g.,][]{Jones&Merrill1976, Ossenkopf+1992, Draine&Lee1984}.  
The optical constants most widely used in astronomy today are likewise based on such composite models.
However, as discussed above, the ability of Fe-rich silicates to reproduce the observed spectra suggests that  
Fe$^{2+}$ cations intrinsic to the silicate lattice may themselves provide significant near-infrared opacity.  
This implies that the ``dirtiness'' of silicates need not arise solely from external inclusions of metallic iron,  
but may also originate from alternative mechanisms inherent to the silicate itself.
While our analysis focuses on near-infrared opacity enhancement, other studies have utilized mid-infrared crystalline features or elemental mapping to infer the presence of Fe-rich silicates in different environments, such as circumstellar dust around AGB stars and presolar silicate grains, based on \citep{Guha-Niyogi+2011, Bose+2012, Floss+2016}.
}

\subsection{Other \rev{near-infrared opacity source} candidates}
\label{subsec:other-conductor}

So far, we have focused on metallic iron as the near-infrared dust opacity source due to its abundance in the atmospheres of G29-38 and GD56. 
However, amorphous carbon and \rev{some forms of iron oxides can exhibit moderate electrical conductivity and} relatively high $k$ values in the near-infrared wavelength range \citep[][and see also the Jena Database of Optical Constants for Cosmic Dust (\url{https://www.astro.uni-jena.de/Laboratory/OCDB/})]{Zubko+1996}.
\rev{Here, we adopt Fe$_3$O$_4$ (magnetite) as a representative example of iron oxides, 
because it exhibits a larger imaginary part of the refractive index ($k$) in the near-infrared 
than  FeO and Fe$_2$O$_3$ (see, e.g., the optical data for FeO, Fe$_2$O$_3$, and Fe$_3$O$_4$ from the Jena database)}\footnote{\rev{Fe$_3$O$_4$ is not simply a mixture of FeO and Fe$_2$O$_3$, but a mixed-valence compound containing both Fe$^{2+}$ and Fe$^{3+}$. 
Intervalence charge transfer between Fe$^{2+}$ and Fe$^{3+}$ ions leads to electron delocalization and metal-like optical behavior, producing high extinction (i.e., large $k$) across the visible to near-infrared wavelengths \citep{Izawa+2019}.
The case of FeO is discussed later in this section.
}}. 
In this section, we discuss the possibility that such materials could be responsible for the near-infrared disk emission around G29-38 when $y=0$. The abundance of carbon and oxygen in the atmosphere of G29-38 is available (Table \ref{tab:ss-elements}).

\begin{figure}
\centering
    \includegraphics[bb=0 0 360 252, width=1.0\columnwidth]{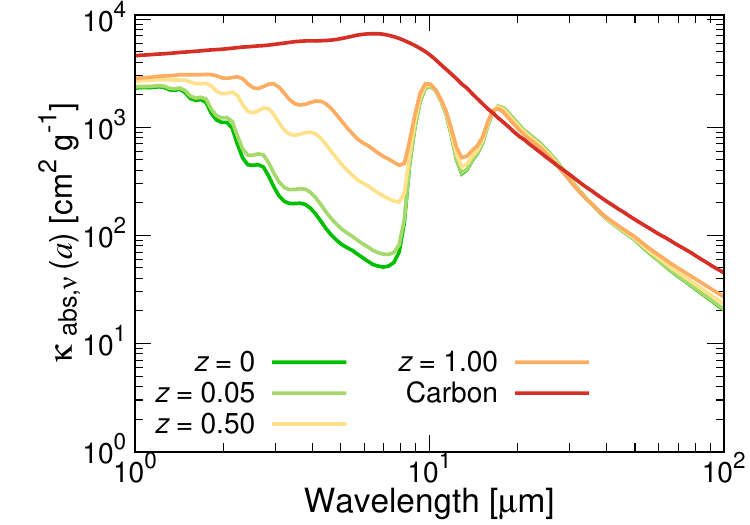}
\caption{Same as Figure~\ref{fig:kappa_mix} but mixed with amorphous carbon with various carbon-to-silicate molar mixing ratios, $z$.
}
\label{fig:kappa_mix-C}
\end{figure}

We calculate the mass absorption opacity of $1~\mu$m-sized MgFeSiO$_4$ dust mixed with amorphous carbon with various carbon-to-silicate molar mixing ratios, $z$ (Equation \ref{eq:y-olivine}), as shown in Figure~\ref{fig:kappa_mix-C}. 
The near-infrared dust opacity is significantly enhanced if $z\gtrsim 0.5$, and the value of $\kappa_{\rm abs}$ at $\lambda\sim 5\ \mu$m is almost the same for dust with $z=0.5$ and dust including metallic iron with $y=1$ (Figure~\ref{fig:kappa_mix}).
Reminding that the dust with $y=1$ well reproduces the observed G29-38 spectrum (Figure \ref{fig:chi_wo_color_G29-38_MgFeSiO4}), the dust with $ z \sim 0.5$ is also expected to reproduce the observed spectrum.

However, such a C/Si value is an order of magnitude higher than the C/Si value inferred from the stellar atmosphere of G29-38, 0.02--0.05 (Table \ref{tab:ss-elements}). 
Therefore, amorphous carbon is unlikely to be the near-infrared emission source more than metallic Fe. 
The discrepancy between C/Si in the \rev{circumstellar} dust and in the stellar atmosphere that arises if we impose the responsibility for the near-infrared emissions on carbon has already been pointed out by \citet{Xu+2014}.

\begin{figure}
\centering
    \includegraphics[bb=0 0 360 252, width=1.0\columnwidth]{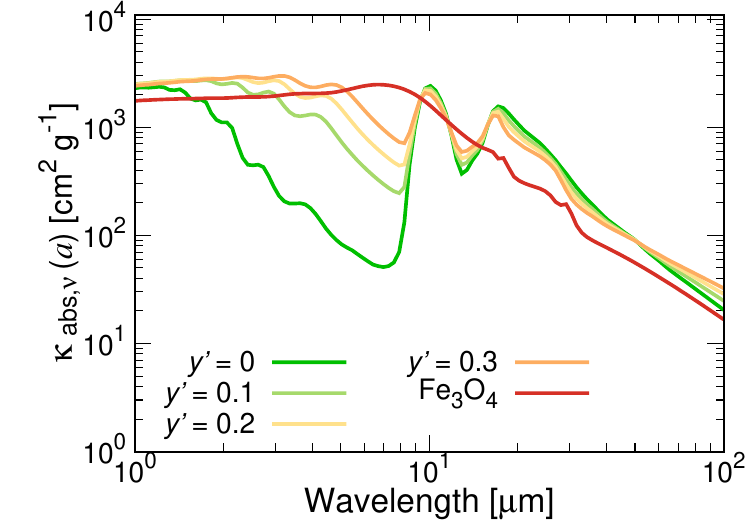}
\caption{Same as Figure~\ref{fig:kappa_mix-C} but mixed with Fe$_3$O$_4$ with various Fe$_3$O$_4$-to-silicate molar mixing ratios, $y'$. }
\label{fig:kappa_mix-FeO}
\end{figure}

We also calculate the opacity of $1~\mu$m-sized MgFeSiO$_4$ dust mixed with Fe$_3$O$_4$ with various Fe$_3$O$_4$-to-silicate molar mixing ratios, $y'$, as shown in Figure~\ref{fig:kappa_mix-FeO}. The refractive index of Fe$_3$O$_4$ is taken from the Jena Database of Optical Constants for Cosmic Dust, \rev{where it is provided as unpublished data}.
When Fe$_3$O$_4$ is mixed in, the near-infrared dust opacity can be increased efficiently even if the mixing ratio is small compared to when the metallic iron is mixed in.
In particular, the value of $\kappa_{\rm abs,\nu}$ at $\lambda\sim 5\ \mu$m is comparable to that for dust with $y'=0.1$ and $y=1$. Therefore, dust with $ y' \sim 0.1$ would possibly reproduce the observed G29-38 spectrum.

In this case, the Fe/Si value of the dust in the disk is $(2-x)+3y' \sim 1.3$, which is consistent with that in the G29-38 atmosphere within the range of the 1$\sigma$ observational uncertainties (Table~\ref{tab:ss-elements}).  
Moreover, the O/Si value of the circumstellar dust is $\sim 4.4$ and matches that in the stellar atmosphere.
Therefore, Fe$_3$O$_4$ can be another solution for the near-infrared emission of G29-38.
Although Fe$_3$O$_4$ exhibits faint features in its opacity at \rev{around 20 and 30~$\mu$m} (Figure \ref{fig:kappa_mix-FeO}), \rev{which likely arise from vibrational modes involiving both Fe$^{2+}$--O and Fe$^{3+}$--O bonds,} it would be challenging to distinguish the emission spectrum of Fe$_3$O$_4$ from that of metallic iron.

\rev{FeO, on the other hand, may also contribute to near-infrared absorption through  electronic transitions within Fe$^{2+}$ ions in the crystal field, 
as discussed in Section~\ref{subsec:fe-rich-silicate}.
These electronic transitions could produce weak spectral features near 1~$\mu$m 
\citep[e.g.,][]{Morris+1985}, which might slightly degrade the spectral fit if prominent.
However, since stellar emission dominates at these wavelengths, such features would likely be diluted 
and have a negligible effect on the observed SED.}

\subsection{Implication of the Presence of Metallic Iron in Dust around White Dwarfs
} \label{subsec:presence-Fe}

The spectral fitting and the elemental abundance inferred from the stellar atmosphere suggest that\rev{, in many cases,} the dust in the disks around G29-38 and GD56 is composed of the amorphous silicate mixed with metallic iron  (see Section \ref{subsec:correlation}).
Such compositions resemble GEMS (glass with embedded metal and sulfide), a major constituent of interplanetary dust particles in the solar system.
GEMS are spherical amorphous silicate particles that contain Fe-Ni metal and sulfides in the form of nano-particle inclusions \citep[e.g.,][]{Bradley+1994}.

\begin{figure}
\centering
\includegraphics[bb=0 0 360 252, width=1.0\columnwidth]{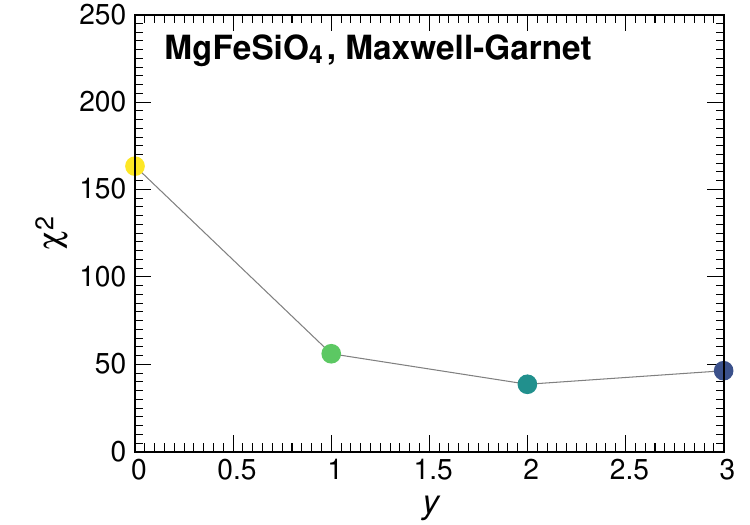}
\caption{Same as Figure \ref{fig:chi_wo_color_G29-38_MgFeSiO4} but adopting the Maxwell--Garnett effective medium theory to calculate the iron-bearing dust opacity.
}
\label{fig:chi_wo_color_G29-38_MgFeSiO4-MG}
\end{figure}

In this study, we have employed the Bruggeman theory to compute the optical properties of dust mixtures. However, the Maxwell--Garnett effective medium theory, which distinguishes between matrix and inclusions  \citep[e.g.,][]{Bohren&Huffman83}, is likely to better apply to GEMS particles.
Therefore, we recalculate the opacity of dust mixed with metallic iron using the Maxwell--Garnett theory and perform the same spectral fitting for G29-38 (Figure \ref{fig:chi_wo_color_G29-38_MgFeSiO4-MG}) as in Section \ref{subsec:Fe-G29-38}.
As with the Bruggeman theory, the addition of metallic iron increases the near-infrared dust opacity, and the iron-bearing dust reproduces the observed G29-38 spectrum better than the iron-free dust.
The best-fit $y$ from the Maxwell--Garnett theory is twice larger than that from the Bruggeman theory.
We note that the Maxwell--Garnett theory is applicable when the volume fraction of inclusions is smaller than that of the matrix, and most of the $y$-values considered in Figure \ref{fig:chi_wo_color_G29-38_MgFeSiO4-MG} satisfy this condition. For example, a metallic iron abundance of $y=2$ corresponds to a volume fraction of 25\%.

If the dust in the disks around G29-38 and GD56 resemble GEMS, its presence could possibly imply that the rapid sublimation and re-condensation repeatedly takes place in these disks, as one of the possible formation processes of the solar system GEMS is rapid condensation from high-temperature gas \citep{Keller+2011,Matsuno+2021,Kim+2021}. 
Rapid condensation may naturally occur in the evolution of silicate dust in disks around white dwarfs \citep{Okuya+2023}. The silicate dust drifts into the silicate sublimation line, producing silicate gas. The gas can diffuse outward and quickly re-condense into the silicate dust outside the silicate sublimation line, which may produce GEMS-analogs.

Inclusion of iron oxides such as Fe$_3$O$_4$ 
can also explain the observed spectrum of G29-38 as discussed in Section \ref{subsec:other-conductor}. 
In contrast, Fe$_3$O$_4$ is absent in the primitive GEMS grains of the solar system \citep{Bradley+1994}, likely due to their formation under reducing conditions.
Condensation experiments for GEMS demonstrate that inclusions in silicates change their composition depending on the redox state, from Fe$_3$O$_4$ to Fe to FeSi \citep{Matsuno+2021}. 
Therefore, if we can identify Fe$_3$O$_4$ in amorphous silicates in a white-dwarf disk, it would suggest that the disk is an oxidizing environment.

While the grain size of GEMS found in the solar system is sub-micron \citep[e.g.,][]{Keller+2011}, 
our best-fit spectral fitting model contains sub-micron to $\sim 10^3\ \mu$m-sized particles (Tables \ref{tab:param-MgFeSiO4-G29} and \ref{tab:param-MgFeSiO4-GD56}).
Therefore, not all the dust particles in the disks around G29-38 and GD56 are similar to GEMS.
In particular, the super-micron-sized particles may be primordial, having not undergone the aforementioned sublimation and re-condensation processes to form GEMS-analogs after their formation through mutual collisions \citep[e.g.,][]{Kenyon+2017}.
In such a case, the abundance and mixing manner of metallic iron in the dust may depend on the particle size. 
\rev{Additionally, laboratory-produced amorphous silicate analogs also exhibit a correlation between metallic iron inclusion size and silicate grain size, although the underlying mechanism is different \citep{Enomoto+2025}.}
We leave incorporating the metallic iron inclusion dependence on the particle size into our emission model for future work.

\section{Conclusion} \label{sec:conclusion}
We have investigated the possibility that conducting materials, such as metallic iron, included in silicate dust are the origin of the near-infrared emission in the observed spectra of the disks around the white dwarfs G29-38 and GD56.
We have analyzed the abundance of metallic iron in the dust and the physical parameters of our disk thermal emission model that can best fit the observed spectra.

We have found that \rev{silicate dust containing metallic iron}
enhances the near-infrared dust opacity (Figure \ref{fig:kappa_mix}) and reproduces the observed G29-38 spectrum better than \rev{silicate dust without metallic iron} (Figures \ref{fig:Fe-dep-G29-38} and \ref{fig:chi_wo_color_G29-38_MgFeSiO4}). 
In particular, dust with a metal-to-silicate molar mixing ratio of approximately unity reproduces both the observed near-infrared excess and 10 $\mu$m silicate feature best.  
This metallic iron abundance is similarly favored \rev{across} different olivine and pyroxene silicate compositions \rev{except for the Fe-rich olivine Mg$_{0.8}$Fe$_{1.2}$SiO$_4$, which reproduces the observed spectrum 
equally well with and without metallic iron.}
Dust with a metal-to-silicate molar mixing ratio from 0 to 2 reproduces the observed GD56 spectrum equally well within the 1-$\sigma$ observational uncertainty (Figure \ref{fig:GD56-silicate}). 
Exceptionally, Mg-pure silicates strongly favor iron-bearing dust.
A re-analysis of future observational data using the JWST MIRI \citep{Kate+2023jwst} and NIRSpec is expected to place tighter constraints on the metallic iron abundance and silicate composition.

We have compared our derived best-fit iron abundance in the circumstellar dust with that in the stellar atmosphere corrected by the sinking timescale for each element.
If these values show an agreement, it suggests that the gas flow accreting onto the stellar surface has a composition similar to the dust in the surrounding disk.
For G29-38, the best-fit Fe/Si values derived from three dust compositions---MgSiO$_3$+Fe, Mg$_{0.7}$Fe$_{0.3}$SiO$_3$+Fe, and Mg$_{0.8}$Fe$_{1.2}$SiO$_4$ with no metallic iron---are consistent with that in the stellar atmosphere (Figure \ref{fig:G29-38_vs_stellar}). For GD56, the iron abundance in the dust is consistent with that in the stellar atmosphere across all silicate compositions considered (Figure \ref{fig:GD56_vs_stellar}).
However, making a more precise comparison is difficult with
the existing stellar atmospheric data because of the large observational uncertainties of the stellar elemental abundances.

\rev{A caveat in our opacity modeling is that  
the optical data of ferro-magnesium silicates
used in this study were derived from samples containing a substantial amount of Fe$^{3+}$ in addition to Fe$^{2+}$. 
If all the iron in the silicate were in the form of Fe$^{2+}$, its characteristic crystal-field transitions could further enhance the near-infrared opacity. 
As a result, the best-fit metal-to-silicate mixing ratio ($y$ value) obtained in our analysis 
might be smaller, yielding a dust composition more consistent with the stellar atmospheric abundance. 
This suggests that the near-infrared emissivity of white dwarf disks 
may not necessarily require metallic iron inclusions but could also arise from intrinsic Fe$^{2+}$ within Fe-rich silicates.}

We have also discussed that other \rev{moderately} conducting materials, such as amorphous carbon and iron oxides (\rev{e.g.,} Fe$_3$O$_4$), could contribute to the observed near-infrared emission in G29-38.
Based on the comparison between their required abundances in the dust to sufficiently increase the near-infrared opacity and those inferred from the stellar atmosphere, the amorphous carbon is unlikely to be the source, whereas \rev{iron oxides} can be a viable alternative.

If \rev{the near-infrared opacity of disk dust originates from metallic iron} embedded in the form of nano-particles within silicate dust, similar to the solar system GEMS grains, their presence may be potential evidence that the rapid sublimation and re-condensation processes repeatedly take places in a white-dwarf disk. 
Such processes are thought to naturally occur in the evolution of silicate dust and gas around a white dwarf \citep{Okuya+2023}.
As the composition of inclusions in condensates reflects redox conditions \citep{Matsuno+2021}, the presence of metallic iron may suggest a reducing environment, whereas Fe$_3$O$_4$ inclusions may indicate an oxidizing environment in the white dwarf disk.

While we have analyzed the spectra of disks around white dwarfs with iron-rich atmospheres, additional samples are required to statistically study the correlation between the iron abundance in circumstellar dust and that in white dwarf atmospheres. In particular, systems with less or no iron in their atmospheres are needed to complement our analysis.
The \rev{ongoing and} upcoming JWST MIRI survey of circumstellar dust will significantly expand the sample of high-precision dust emission spectra. 
Furthermore, spectroscopic surveys of elemental abundances in white dwarf atmospheres by SDSS, DESI, and WEAVE will contribute to increasing the number of white dwarfs with detailed abundance measurements.

\section*{Acknowledgments} 
\begin{acknowledgments}
\rev{The authors thank the referee for constructive comments that greatly improved the discussion on mineralogical interpretations, and
}
Hiroshi Kobayashi, Siyi Xu, and Akemi Tamanai for useful discussions. 
This work was supported by JSPS KAKENHI Grant Numbers JP22KJ3094 and JP23K25923. 
This research has made use of the SIMBAD database, CDS, Strasbourg Astronomical Observatory, France.
\end{acknowledgments}

%

\vspace{5mm}
\facilities{Spitzer(IRS), Spitzer(IRAC), CTIO:2MASS, FLWO:2MASS}


\software{OpTool \citep{Dominik+2021}
          }



\appendix

\section{Comparison with previous disk models and parameters} \label{sec:comparison}
We comment on the differences between our disk model and parameters in Section \ref{subsec:disk-model} and those used in previous studies.
Note that our model does not rule out the validity of the two-radial-component model because we cannot uniquely determine which disk structure is correct with the existing spatially unresolved observational data.

The most commonly used model for analyzing observed spectra, including silicate features, is the two-radial-component disk model \citep{Jura+2007AJ, Reach+2009, Jura+2009AJ, Xu+2018}. This model consists of a geometrically flat, optically thick inner region and an optically thin outer region. The near-infrared emission is primarily produced by the optically thick region, while the mid-infrared silicate features arise from the optically thin outer region. The contribution ratio between these regions is mainly controlled by the disk inclination.

In contrast, our disk emission model assumes a single radial region with two vertical layers: an optically thin surface layer and a cooler disk interior (see Section \ref{subsec:disk-model}). Unlike the two-radial-component model, which requires an inner opaque disk, our model reproduces the near-infrared emission by varying the metal-to-silicate mixing ratio in the dust.

Our model may calculate thermal emission spectra more self-consistently than the two-radial-component model. For instance, the optical thickness of the disk to its own emission depends directly on the disk physical parameters in our model, while the two-radial-component model divides optically thick and thin regions solely based on radial distance. 
Possibly due to this, our best-fit disk physical parameters for G29-38 align with those constrained by the most recent radiative transfer simulations using RADMC-3 \citep{Ballering+2022}. Their model also suggests a vertically extended disk with $\theta_{\rm hoa} > 1.4^{\circ}$, a ring-like single region of $\Delta r \sim 10 R_{\star}$, and an inner edge located at $\sim 100 R_{\star}$ (see also Table \ref{tab:param-MgFeSiO4-G29} for the comparison).

\bibliographystyle{aasjournal}

\begin{thebibliography}{99}
\expandafter\ifx\csname natexlab\endcsname\relax\def\natexlab#1{#1}\fi
\providecommand{\url}[1]{\href{#1}{#1}}
\providecommand{\dodoi}[1]{doi:~\href{http://doi.org/#1}{\nolinkurl{#1}}}
\providecommand{\doeprint}[1]{\href{http://ascl.net/#1}{\nolinkurl{http://ascl.net/#1}}}
\providecommand{\doarXiv}[1]{\href{https://arxiv.org/abs/#1}{\nolinkurl{https://arxiv.org/abs/#1}}}

\bibitem[{{Ballering} {et~al.}(2022){Ballering}, {Levens}, {Su}, \&
  {Cleeves}}]{Ballering+2022}
{Ballering}, N.~P., {Levens}, C.~I., {Su}, K. Y.~L., \& {Cleeves}, L.~I. 2022,
  \apj, 939, 108, \dodoi{10.3847/1538-4357/ac9a4a}

\bibitem[{{Bohren} \& {Huffman}(1983)}]{Bohren&Huffman83}
{Bohren}, C.~F., \& {Huffman}, D.~R. 1983, {Absorption and scattering of light
  by small particles}

\bibitem[{Bose {et~al.}(2012)Bose, Floss, Stadermann, Stroud, \&
  Speck}]{Bose+2012}
Bose, M., Floss, C., Stadermann, F.~J., Stroud, R.~M., \& Speck, A.~K. 2012,
  Geochimica et Cosmochimica Acta, 93, 77,
  \dodoi{https://doi.org/10.1016/j.gca.2012.06.027}

\bibitem[{{Bradley}(1994)}]{Bradley+1994}
{Bradley}, J.~P. 1994, \gca, 58, 2123, \dodoi{10.1016/0016-7037(94)90290-9}

\bibitem[{{Brouwers} {et~al.}(2022){Brouwers}, {Bonsor}, \&
  {Malamud}}]{Brouwers+2022}
{Brouwers}, M.~G., {Bonsor}, A., \& {Malamud}, U. 2022, \mnras, 509, 2404,
  \dodoi{10.1093/mnras/stab3009}

\bibitem[{{Burns}(1993)}]{Burns1993}
{Burns}, R.~G. 1993, {Mineralogical Applications of Crystal Field Theory}

\bibitem[{{Chiang} \& {Goldreich}(1997)}]{Chiang+1997}
{Chiang}, E.~I., \& {Goldreich}, P. 1997, \apj, 490, 368,
  \dodoi{10.1086/304869}

\bibitem[{{Chiang} {et~al.}(2001){Chiang}, {Joung}, {Creech-Eakman}, {Qi},
  {Kessler}, {Blake}, \& {van Dishoeck}}]{Chiang+2001}
{Chiang}, E.~I., {Joung}, M.~K., {Creech-Eakman}, M.~J., {et~al.} 2001, \apj,
  547, 1077, \dodoi{10.1086/318427}

\bibitem[{{Cutri} {et~al.}(2003){Cutri}, {Skrutskie}, {van Dyk}, {Beichman},
  {Carpenter}, {Chester}, {Cambresy}, {Evans}, {Fowler}, {Gizis}, {Howard},
  {Huchra}, {Jarrett}, {Kopan}, {Kirkpatrick}, {Light}, {Marsh}, {McCallon},
  {Schneider}, {Stiening}, {Sykes}, {Weinberg}, {Wheaton}, {Wheelock}, \&
  {Zacarias}}]{Cutri+2003}
{Cutri}, R.~M., {Skrutskie}, M.~F., {van Dyk}, S., {et~al.} 2003, {VizieR
  Online Data Catalog: 2MASS All-Sky Catalog of Point Sources (Cutri+ 2003)},
  VizieR On-line Data Catalog: II/246. Originally published in: University of
  Massachusetts and Infrared Processing and Analysis Center, (IPAC/California
  Institute of Technology) (2003)

\bibitem[{{Dominik} {et~al.}(2021){Dominik}, {Min}, \& {Tazaki}}]{Dominik+2021}
{Dominik}, C., {Min}, M., \& {Tazaki}, R. 2021, {OpTool: Command-line driven
  tool for creating complex dust opacities}, Astrophysics Source Code Library,
  record ascl:2104.010

\bibitem[{{Dorschner} {et~al.}(1995){Dorschner}, {Begemann}, {Henning},
  {Jaeger}, \& {Mutschke}}]{Dorschner+1995}
{Dorschner}, J., {Begemann}, B., {Henning}, T., {Jaeger}, C., \& {Mutschke}, H.
  1995, \aap, 300, 503

\bibitem[{{Draine} \& {Lee}(1984)}]{Draine&Lee1984}
{Draine}, B.~T., \& {Lee}, H.~M. 1984, \apj, 285, 89, \dodoi{10.1086/162480}

\bibitem[{Enomoto {et~al.}(2025)Enomoto, Takigawa, Chihara, \&
  Koike}]{Enomoto+2025}
Enomoto, H., Takigawa, A., Chihara, H., \& Koike, C. 2025, Hidden Metallic Iron
  in Amorphous Silicate Dust? Insights from Condensation Experiments and
  Mid-Infrared Spectroscopy, \dodoi{10.48550/arXiv.2511.16217}

\bibitem[{{Farihi}(2016)}]{Farihi2016}
{Farihi}, J. 2016, \nar, 71, 9, \dodoi{10.1016/j.newar.2016.03.001}

\bibitem[{{Farihi} {et~al.}(2025){Farihi}, {Su}, {Melis}, {Kenyon}, {Swan},
  {Redfield}, {Wyatt}, \& {Debes}}]{Farihi+2025}
{Farihi}, J., {Su}, K.~Y.~L., {Melis}, C., {et~al.} 2025, \apjl, 981, L5,
  \dodoi{10.3847/2041-8213/adae88}

\bibitem[{{Farihi} {et~al.}(2018){Farihi}, {van Lieshout}, {Cauley}, {Dennihy},
  {Su}, {Kenyon}, {Wilson}, {Toloza}, {G{\"a}nsicke}, {von Hippel}, {Redfield},
  {Debes}, {Xu}, {Rogers}, {Bonsor}, {Swan}, {Pala}, \& {Reach}}]{Farihi+2018}
{Farihi}, J., {van Lieshout}, R., {Cauley}, P.~W., {et~al.} 2018, \mnras, 481,
  2601, \dodoi{10.1093/mnras/sty2331}

\bibitem[{Floss \& Haenecour(2016)}]{Floss+2016}
Floss, C., \& Haenecour, P. 2016, GEOCHEMICAL JOURNAL, 50, 3,
  \dodoi{10.2343/geochemj.2.0377}

\bibitem[{{Girven} {et~al.}(2012){Girven}, {Brinkworth}, {Farihi},
  {G{\"a}nsicke}, {Hoard}, {Marsh}, \& {Koester}}]{Girven+2012}
{Girven}, J., {Brinkworth}, C.~S., {Farihi}, J., {et~al.} 2012, \apj, 749, 154,
  \dodoi{10.1088/0004-637X/749/2/154}

\bibitem[{{Guha Niyogi} {et~al.}(2011){Guha Niyogi}, {Speck}, \&
  {Onaka}}]{Guha-Niyogi+2011}
{Guha Niyogi}, S., {Speck}, A.~K., \& {Onaka}, T. 2011, \apj, 733, 93,
  \dodoi{10.1088/0004-637X/733/2/93}

\bibitem[{{Hollands} {et~al.}(2017){Hollands}, {Koester}, {Alekseev},
  {Herbert}, \& {G{\"a}nsicke}}]{Hollands+2017}
{Hollands}, M.~A., {Koester}, D., {Alekseev}, V., {Herbert}, E.~L., \&
  {G{\"a}nsicke}, B.~T. 2017, \mnras, 467, 4970, \dodoi{10.1093/mnras/stx250}

\bibitem[{Izawa {et~al.}(2019)Izawa, Cloutis, Rhind, Mertzman, Applin,
  Stromberg, \& Sherman}]{Izawa+2019}
Izawa, M.~R., Cloutis, E.~A., Rhind, T., {et~al.} 2019, Icarus, 319, 525,
  \dodoi{https://doi.org/10.1016/j.icarus.2018.10.002}

\bibitem[{{J{\"a}ger} {et~al.}(2003){J{\"a}ger}, {Dorschner}, {Mutschke},
  {Posch}, \& {Henning}}]{Jager+2003}
{J{\"a}ger}, C., {Dorschner}, J., {Mutschke}, H., {Posch}, T., \& {Henning}, T.
  2003, \aap, 408, 193, \dodoi{10.1051/0004-6361:20030916}

\bibitem[{{Jones} \& {Merrill}(1976)}]{Jones&Merrill1976}
{Jones}, T.~W., \& {Merrill}, K.~M. 1976, \apj, 209, 509,
  \dodoi{10.1086/154746}

\bibitem[{{Jura}(2003)}]{Jura2003}
{Jura}, M. 2003, \apjl, 584, L91, \dodoi{10.1086/374036}

\bibitem[{{Jura} {et~al.}(2007{\natexlab{a}}){Jura}, {Farihi}, \&
  {Zuckerman}}]{Jura+2007}
{Jura}, M., {Farihi}, J., \& {Zuckerman}, B. 2007{\natexlab{a}}, \apj, 663,
  1285, \dodoi{10.1086/518767}

\bibitem[{{Jura} {et~al.}(2009){Jura}, {Farihi}, \& {Zuckerman}}]{Jura+2009AJ}
---. 2009, \aj, 137, 3191, \dodoi{10.1088/0004-6256/137/2/3191}

\bibitem[{{Jura} {et~al.}(2007{\natexlab{b}}){Jura}, {Farihi}, {Zuckerman}, \&
  {Becklin}}]{Jura+2007AJ}
{Jura}, M., {Farihi}, J., {Zuckerman}, B., \& {Becklin}, E.~E.
  2007{\natexlab{b}}, \aj, 133, 1927, \dodoi{10.1086/512734}

\bibitem[{{Jura} \& {Young}(2014)}]{Jura&Young2014}
{Jura}, M., \& {Young}, E.~D. 2014, Annual Review of Earth and Planetary
  Sciences, 42, 45, \dodoi{10.1146/annurev-earth-060313-054740}

\bibitem[{{Keller} \& {Messenger}(2011)}]{Keller+2011}
{Keller}, L.~P., \& {Messenger}, S. 2011, \gca, 75, 5336,
  \dodoi{10.1016/j.gca.2011.06.040}

\bibitem[{{Kemper} {et~al.}(2002){Kemper}, {de Koter}, {Waters}, {Bouwman}, \&
  {Tielens}}]{Kemper+2002}
{Kemper}, F., {de Koter}, A., {Waters}, L.~B.~F.~M., {Bouwman}, J., \&
  {Tielens}, A.~G.~G.~M. 2002, \aap, 384, 585,
  \dodoi{10.1051/0004-6361:20020036}

\bibitem[{{Kenyon} \& {Bromley}(2017)}]{Kenyon+2017}
{Kenyon}, S.~J., \& {Bromley}, B.~C. 2017, \apj, 850, 50,
  \dodoi{10.3847/1538-4357/aa9570}

\bibitem[{{Kilic} {et~al.}(2006){Kilic}, {von Hippel}, {Leggett}, \&
  {Winget}}]{Kilic+2006}
{Kilic}, M., {von Hippel}, T., {Leggett}, S.~K., \& {Winget}, D.~E. 2006, \apj,
  646, 474, \dodoi{10.1086/504682}

\bibitem[{{Kim} {et~al.}(2021){Kim}, {Takigawa}, {Tsuchiyama}, {Matsuno},
  {Enju}, {Kawano}, \& {Komaki}}]{Kim+2021}
{Kim}, T.~H., {Takigawa}, A., {Tsuchiyama}, A., {et~al.} 2021, \aap, 656, A42,
  \dodoi{10.1051/0004-6361/202141216}

\bibitem[{{Koester}(2009)}]{Koester2009}
{Koester}, D. 2009, \aap, 498, 517, \dodoi{10.1051/0004-6361/200811468}

\bibitem[{{Koester} {et~al.}(2014){Koester}, {G{\"a}nsicke}, \&
  {Farihi}}]{Koester+2014}
{Koester}, D., {G{\"a}nsicke}, B.~T., \& {Farihi}, J. 2014, \aap, 566, A34,
  \dodoi{10.1051/0004-6361/201423691}

\bibitem[{{Koester} \& {Wilken}(2006)}]{Koester&Wilken2006}
{Koester}, D., \& {Wilken}, D. 2006, \aap, 453, 1051,
  \dodoi{10.1051/0004-6361:20064843}

\bibitem[{{Kusaka} {et~al.}(1970){Kusaka}, {Nakano}, \&
  {Hayashi}}]{Kusaka+1970}
{Kusaka}, T., {Nakano}, T., \& {Hayashi}, C. 1970, Progress of Theoretical
  Physics, 44, 1580, \dodoi{10.1143/PTP.44.1580}

\bibitem[{{Lebouteiller} {et~al.}(2011){Lebouteiller}, {Barry}, {Spoon},
  {Bernard-Salas}, {Sloan}, {Houck}, \& {Weedman}}]{Lebouteiller+2011}
{Lebouteiller}, V., {Barry}, D.~J., {Spoon}, H.~W.~W., {et~al.} 2011, \apjs,
  196, 8, \dodoi{10.1088/0067-0049/196/1/8}

\bibitem[{{Li} {et~al.}(2021){Li}, {Mustill}, \& {Davies}}]{Li+2021}
{Li}, D., {Mustill}, A.~J., \& {Davies}, M.~B. 2021, \mnras, 508, 5671,
  \dodoi{10.1093/mnras/stab2949}

\bibitem[{{Malamud} \& {Perets}(2020{\natexlab{a}})}]{Malamud+2020a}
{Malamud}, U., \& {Perets}, H.~B. 2020{\natexlab{a}}, \mnras, 492, 5561,
  \dodoi{10.1093/mnras/staa142}

\bibitem[{{Malamud} \& {Perets}(2020{\natexlab{b}})}]{Malamud+2020b}
---. 2020{\natexlab{b}}, \mnras, 493, 698, \dodoi{10.1093/mnras/staa143}

\bibitem[{{Matsuno} {et~al.}(2021){Matsuno}, {Tsuchiyama}, {Watanabe},
  {Tanaka}, {Takigawa}, {Enju}, {Koike}, {Chihara}, \& {Miyake}}]{Matsuno+2021}
{Matsuno}, J., {Tsuchiyama}, A., {Watanabe}, T., {et~al.} 2021, \apj, 911, 47,
  \dodoi{10.3847/1538-4357/abe5a0}

\bibitem[{{Miyake} \& {Nakagawa}(1993)}]{MiyakeNakagawa1993}
{Miyake}, K., \& {Nakagawa}, Y. 1993, \icarus, 106, 20,
  \dodoi{10.1006/icar.1993.1156}

\bibitem[{Morris {et~al.}(1985)Morris, Lauer~Jr., Lawson, Gibson~Jr., Nace, \&
  Stewart}]{Morris+1985}
Morris, R.~V., Lauer~Jr., H.~V., Lawson, C.~A., {et~al.} 1985, Journal of
  Geophysical Research: Solid Earth, 90, 3126,
  \dodoi{https://doi.org/10.1029/JB090iB04p03126}

\bibitem[{{Okuya} {et~al.}(2023){Okuya}, {Ida}, {Hyodo}, \&
  {Okuzumi}}]{Okuya+2023}
{Okuya}, A., {Ida}, S., {Hyodo}, R., \& {Okuzumi}, S. 2023, \mnras, 519, 1657,
  \dodoi{10.1093/mnras/stac3522}

\bibitem[{{Ordal} {et~al.}(1988){Ordal}, {Bell}, {Alexander}, {Newquist}, \&
  {Querry}}]{Ordal+1988}
{Ordal}, M.~A., {Bell}, R.~J., {Alexander}, Jr., R.~W., {Newquist}, L.~A., \&
  {Querry}, M.~R. 1988, \ao, 27, 1203, \dodoi{10.1364/AO.27.001203}

\bibitem[{{Ossenkopf} {et~al.}(1992){Ossenkopf}, {Henning}, \&
  {Mathis}}]{Ossenkopf+1992}
{Ossenkopf}, V., {Henning}, T., \& {Mathis}, J.~S. 1992, \aap, 261, 567

\bibitem[{{Paquette} {et~al.}(1986){Paquette}, {Pelletier}, {Fontaine}, \&
  {Michaud}}]{Paquette1986}
{Paquette}, C., {Pelletier}, C., {Fontaine}, G., \& {Michaud}, G. 1986, \apjs,
  61, 177, \dodoi{10.1086/191111}

\bibitem[{{Rafikov}(2011)}]{Rafikov2011}
{Rafikov}, R.~R. 2011, \apjl, 732, L3, \dodoi{10.1088/2041-8205/732/1/L3}

\bibitem[{{Rafikov} \& {Garmilla}(2012)}]{Rafikov+2012}
{Rafikov}, R.~R., \& {Garmilla}, J.~A. 2012, \apj, 760, 123,
  \dodoi{10.1088/0004-637X/760/2/123}

\bibitem[{{Reach} {et~al.}(2005){Reach}, {Kuchner}, {von Hippel}, {Burrows},
  {Mullally}, {Kilic}, \& {Winget}}]{Reach+2005}
{Reach}, W.~T., {Kuchner}, M.~J., {von Hippel}, T., {et~al.} 2005, \apjl, 635,
  L161, \dodoi{10.1086/499561}

\bibitem[{{Reach} {et~al.}(2009){Reach}, {Lisse}, {von Hippel}, \&
  {Mullally}}]{Reach+2009}
{Reach}, W.~T., {Lisse}, C., {von Hippel}, T., \& {Mullally}, F. 2009, \apj,
  693, 697, \dodoi{10.1088/0004-637X/693/1/697}

\bibitem[{{Reach} {et~al.}(2025){Reach}, {Kilic}, {Lisse}, {Debes}, {von
  Hippel}, {Azartash-Namin}, {Albert}, {Mullally}, {Mullally}, {Cracraft},
  {Bernice}, \& {Erickson}}]{Reach+2025}
{Reach}, W.~T., {Kilic}, M., {Lisse}, C.~M., {et~al.} 2025, arXiv e-prints,
  arXiv:2510.07595, \dodoi{10.48550/arXiv.2510.07595}

\bibitem[{{Rocchetto} {et~al.}(2015){Rocchetto}, {Farihi}, {G{\"a}nsicke}, \&
  {Bergfors}}]{Rocchetto+2015}
{Rocchetto}, M., {Farihi}, J., {G{\"a}nsicke}, B.~T., \& {Bergfors}, C. 2015,
  \mnras, 449, 574, \dodoi{10.1093/mnras/stv282}

\bibitem[{{Ruden} \& {Pollack}(1991)}]{Ruden&Pollack1991}
{Ruden}, S.~P., \& {Pollack}, J.~B. 1991, \apj, 375, 740,
  \dodoi{10.1086/170239}

\bibitem[{{Speck} {et~al.}(2015){Speck}, {Pitman}, \&
  {Hofmeister}}]{Speck+2015}
{Speck}, A.~K., {Pitman}, K.~M., \& {Hofmeister}, A.~M. 2015, \apj, 809, 65,
  \dodoi{10.1088/0004-637X/809/1/65}

\bibitem[{{Speck} {et~al.}(2011){Speck}, {Whittington}, \&
  {Hofmeister}}]{Speck+2011}
{Speck}, A.~K., {Whittington}, A.~G., \& {Hofmeister}, A.~M. 2011, \apj, 740,
  93, \dodoi{10.1088/0004-637X/740/2/93}

\bibitem[{{Su} {et~al.}(2023){Su}, {Ballering}, {Bonsor}, {Rogers}, \&
  {Xu}}]{Kate+2023jwst}
{Su}, K. Y.~L., {Ballering}, N., {Bonsor}, A., {Rogers}, L., \& {Xu}, S. 2023,
  {Characterizing the End Stage of Exoplanetary Systems}, JWST Proposal. Cycle
  2, ID. \#3271

\bibitem[{{Subasavage} {et~al.}(2017){Subasavage}, {Jao}, {Henry}, {Harris},
  {Dahn}, {Bergeron}, {Dufour}, {Dunlap}, {Barlow}, {Ianna}, {L{\'e}pine}, \&
  {Margheim}}]{Subasavage+2017}
{Subasavage}, J.~P., {Jao}, W.-C., {Henry}, T.~J., {et~al.} 2017, \aj, 154, 32,
  \dodoi{10.3847/1538-3881/aa76e0}

\bibitem[{{Swan} {et~al.}(2024){Swan}, {Farihi}, {Su}, \& {Desch}}]{Swan+2024}
{Swan}, A., {Farihi}, J., {Su}, K. Y.~L., \& {Desch}, S.~J. 2024, \mnras, 529,
  L41, \dodoi{10.1093/mnrasl/slad198}

\bibitem[{{Veras}(2021)}]{Veras2021}
{Veras}, D. 2021, in Oxford Research Encyclopedia of Planetary Science, 1,
  \dodoi{10.1093/acrefore/9780190647926.013.238}

\bibitem[{{Veras} {et~al.}(2014){Veras}, {Leinhardt}, {Bonsor}, \&
  {G{\"a}nsicke}}]{Veras+2014}
{Veras}, D., {Leinhardt}, Z.~M., {Bonsor}, A., \& {G{\"a}nsicke}, B.~T. 2014,
  \mnras, 445, 2244, \dodoi{10.1093/mnras/stu1871}

\bibitem[{{Veras} {et~al.}(2015){Veras}, {Leinhardt}, {Eggl}, \&
  {G{\"a}nsicke}}]{Veras+2015}
{Veras}, D., {Leinhardt}, Z.~M., {Eggl}, S., \& {G{\"a}nsicke}, B.~T. 2015,
  \mnras, 451, 3453, \dodoi{10.1093/mnras/stv1195}

\bibitem[{{Wenger} {et~al.}(2000){Wenger}, {Ochsenbein}, {Egret}, {Dubois},
  {Bonnarel}, {Borde}, {Genova}, {Jasniewicz}, {Lalo{\"e}}, {Lesteven}, \&
  {Monier}}]{SIMBAD}
{Wenger}, M., {Ochsenbein}, F., {Egret}, D., {et~al.} 2000, \aaps, 143, 9,
  \dodoi{10.1051/aas:2000332}

\bibitem[{{Xu} {et~al.}(2019){Xu}, {Dufour}, {Klein}, {Melis}, {Monson},
  {Zuckerman}, {Young}, \& {Jura}}]{Xu+2019}
{Xu}, S., {Dufour}, P., {Klein}, B., {et~al.} 2019, \aj, 158, 242,
  \dodoi{10.3847/1538-3881/ab4cee}

\bibitem[{{Xu} {et~al.}(2014){Xu}, {Jura}, {Koester}, {Klein}, \&
  {Zuckerman}}]{Xu+2014}
{Xu}, S., {Jura}, M., {Koester}, D., {Klein}, B., \& {Zuckerman}, B. 2014,
  \apj, 783, 79, \dodoi{10.1088/0004-637X/783/2/79}

\bibitem[{{Xu} {et~al.}(2018){Xu}, {Su}, {Rogers}, {Bonsor}, {Olofsson},
  {Veras}, {van Lieshout}, {Dufour}, {Green}, {Schlawin}, {Farihi}, {Wilson},
  {Wilson}, \& {G{\"a}nsicke}}]{Xu+2018}
{Xu}, S., {Su}, K. Y.~L., {Rogers}, L.~K., {et~al.} 2018, \apj, 866, 108,
  \dodoi{10.3847/1538-4357/aadcfe}

\bibitem[{{Zeidler} {et~al.}(2011){Zeidler}, {Posch}, {Mutschke}, {Richter}, \&
  {Wehrhan}}]{Zeidler+2011}
{Zeidler}, S., {Posch}, T., {Mutschke}, H., {Richter}, H., \& {Wehrhan}, O.
  2011, \aap, 526, A68, \dodoi{10.1051/0004-6361/201015219}

\bibitem[{{Zubko} {et~al.}(1996){Zubko}, {Mennella}, {Colangeli}, \&
  {Bussoletti}}]{Zubko+1996}
{Zubko}, V.~G., {Mennella}, V., {Colangeli}, L., \& {Bussoletti}, E. 1996,
  \mnras, 282, 1321, \dodoi{10.1093/mnras/282.4.1321}

\bibitem[{{Zuckerman} \& {Becklin}(1987)}]{Zuckerman+1987}
{Zuckerman}, B., \& {Becklin}, E.~E. 1987, \nat, 330, 138,
  \dodoi{10.1038/330138a0}

\bibitem[{{Zuckerman} {et~al.}(2003){Zuckerman}, {Koester}, {Reid}, \&
  {H{\"u}nsch}}]{Zuckerman+2003}
{Zuckerman}, B., {Koester}, D., {Reid}, I.~N., \& {H{\"u}nsch}, M. 2003, \apj,
  596, 477, \dodoi{10.1086/377492}

\bibitem[{{Zuckerman} {et~al.}(2010){Zuckerman}, {Melis}, {Klein}, {Koester},
  \& {Jura}}]{Zuckerman+2010}
{Zuckerman}, B., {Melis}, C., {Klein}, B., {Koester}, D., \& {Jura}, M. 2010,
  \apj, 722, 725, \dodoi{10.1088/0004-637X/722/1/725}

\end{thebibliography}



\end{document}